%% file: main.tex
\documentclass[letterpaper,twocolumn,10pt]{article}
\usepackage{usenix}
\usepackage{xspace}
\usepackage{epsfig}
\usepackage{caption}
\usepackage{subcaption}
\usepackage{amssymb}
\usepackage{amsmath}
\usepackage{graphicx}
\usepackage{algorithm}
\usepackage[sort]{cite}
\usepackage{hyperref}
\usepackage{breakurl}
\usepackage{verbatim}
\usepackage{boxedminipage}
\usepackage[compact]{titlesec}
\usepackage{paralist}

\long\def\com#1{}

\long\def\xxx#1{}

\long\def\abbr#1#2{#2}			

\date{}

\com{ 
\renewenvironment{itemize}{
   \begin{list}{\labelitemi}{
     \setlength{\topsep}{0.5ex}
     \setlength{\itemsep}{-0pt}
     \setlength{\itemindent}{0pt}
     \setlength{\leftmargin}{\labelwidth}
     \addtolength{\leftmargin}{-8pt}}
}{\end{list}}
}

\defaultleftmargin{\parindent}{}{}{}

\makeatletter
\g@addto@macro \normalsize {%
 \setlength\abovedisplayskip{6pt plus 2pt minus 2pt}%
 \setlength\belowdisplayskip{6pt plus 2pt minus 2pt}%
}
\makeatother

\newcommand{\app}{Verdict\xspace}
\newcommand{\App}{Verdict\xspace}
\title{\Large\bf	
	Proactively Accountable Anonymous Messaging in \App
    \abbr{}{\\{\em Extended Version}}
}

\author{{\rm Henry Corrigan-Gibbs,
  David Isaac Wolinsky, and Bryan Ford} \\
	Yale University
}

\begin{document}

\maketitle

\input{abs}

\input{intro}
\input{bg}

\input{arch}

\input{proto}

\input{crypto}

\input{eval}

\input{rel}

\input{concl}


{\small
\bibliographystyle{plain}
\bibliography{theory,net,soc,sec,os}
}

\abbr{}{
\appendix
\input{malkeys}
\input{protosec}

\input{proof}

\input{zkp}

\input{lazy}

}

\end{document}

%% file: abs.tex
\subsection*{Abstract} 

Among anonymity systems, DC-nets have long held attraction
for their resistance to traffic analysis attacks,
but practical implementations remain vulnerable to internal disruption
or ``jamming'' attacks, which require 
time-consuming detection procedures to resolve.
We present \app, the first practical anonymous group communication system
built using {\em proactively verifiable} DC-nets:
participants use public-key cryptography to construct DC-net ciphertexts,
and use zero-knowledge proofs of knowledge 
to detect and exclude misbehavior {\em before} disruption.
We compare three alternative constructions for verifiable DC-nets:
one using bilinear maps and two based on simpler ElGamal encryption.
While verifiable DC-nets incur higher computational overheads
due to the public-key cryptography involved,
our experiments suggest that Verdict is practical
for anonymous group messaging or microblogging applications,
supporting groups of 100 clients at 1 second per round
or 1000 clients at 10 seconds per round.
Furthermore, we show how existing symmetric-key DC-nets can ``fall back''
to a verifiable DC-net to quickly identify misbehavior,
speeding up previous detections schemes by two orders of magnitude.

%% file: intro.tex
\section{Introduction}
A right to anonymity is fundamental to
democratic culture, freedom of speech~\cite{
balkin04digital,teich99anonymous},
peaceful resistance to repression~\cite{preston11facebook},
and protecting minority rights~\cite{stein03queers}.
Anonymizing relay tools, such as Tor~\cite{dingledine04tor},
offer practical and scalable anonymous communication
but are vulnerable to traffic analysis attacks~\cite{
	bauer07low,murdoch05lowcost,panchenko11website}
feasible for powerful adversaries,
such as ISPs in authoritarian states.

{\em Dining cryptographers networks}~\cite{chaum88dining} ({\em DC-nets})
promise security even against traffic analysis attacks,
and recent systems
such as Herbivore~\cite{goel03herbivore,sirer04eluding}
and Dissent~\cite{corrigangibbs10dissent,wolinsky12dissent}
have improved the scalability of DC-net-style systems.
However, these systems are still vulnerable to internal {\em disruption} attacks
in which a misbehaving member anonymously ``jams'' communication,
either completely or selectively.
Dissent includes a {\em retrospective blame} procedure
that can eventually exclude disruptors,
but at high cost:
tracing a disruptor in a 1,000-member group
takes over 60 minutes~\cite{wolinsky12dissent},
and the protocol makes no communication progress
until it restarts ``from scratch.''
An adversary who infiltrates such a group
with $f$ colluding members can ``sacrifice'' them one at a time
to disrupt {\em all} communication for $f$ contiguous hours
at any time---%
long enough time to cause a communications blackout
before or during an important mass protest, for example.

\App, a novel but practical group anonymity system,
thwarts such disruptions while maintaining
DC-nets' resistance to traffic analysis. 
At \app's core lies a {\em verifiable} DC-net primitive,
derived from theoretical work proposed and formalized
by Golle and Juels~\cite{golle04dining},
which requires participating nodes to prove
{\em proactively} the well-formedness of messages they send. 
The first working system we are aware of
to implement verifiable DC-nets,
\app supports three alternative schemes for comparison:
a pairing scheme using bilinear maps similar to the Golle-Juels approach,
and two schemes based on ElGamal encryption in conventional
integer or elliptic curve groups.
\app incorporates this verifiable core into a client/server architecture
like Dissent's~\cite{wolinsky12dissent},
to achieve scalability and robustness to client churn.
As in Dissent, \app maintains security as long
as {\em at least one} of the participating servers is honest,
and participants need not know or guess which servers are honest.

Due to their reliance on public-key cryptography,
verifiable DC-nets incur higher computation overheads
than traditional DC-nets, which primarily use 
symmetric-key cryptography (e.g., AES).
We expect this CPU cost to be acceptable
in applications where messages are usually short (e.g., chat or microblogging),
where costs are dominated by network delays,
or in groups with relatively open or antagonistic membership
where disruption risks may be high.
Under realistic conditions, we find that \app
can support groups of 100 users while maintaining 1-second messaging latencies,
or 1000-user groups with 10-second latencies.
In a trace-driven evaluation of full-system performance
for a microblogging application,
\app is able to keep up with symmetric-key DC-nets in groups of
up to 250 active users.

In contrast with the above ``purist'' approach, which uses
expensive public-key cryptography to construct {\em all} DC-net ciphertexts,
\app also implements and evaluates a {\em hybrid} approach
that uses symmetric-key DC-nets for data communication
when not under disruption attack,
but leverages verifiable DC-nets to enable the system to respond
much more quickly and inexpensively to disruption attacks.
Dissent uses
a {\em verifiable shuffle}~\cite{neff01verifiable}
to broadcast an {\em accusation} anonymously;
this shuffle dominates the cost of identifying disruptors.
By replacing this verifiable shuffle with a verifiable DC-nets round,
\app preserves the disruption-free performance of symmetric-key DC-nets,
but reduces the time to identify a disruptor in a 1000-node group
by two orders of magnitude, from 20 minutes to 26 seconds.

This paper's primary contributions are:
\begin{compactitem}
\item	the first working implementation and experimental evaluation
	of verifiable DC-nets
	in a practical anonymous communication system,
\item	two novel verifiable DC-nets constructions using
	standard modular integer and elliptic curve groups,
	offering an order of magnitude lower computational cost
	than the original pairing approach~\cite{golle04dining},
\item	a hybrid system design that preserves
  performance of symmetric-key DC-nets,
	while reducing disruption resolution costs by two orders of magnitude, and
\item	experimental evidence suggesting that verifiable DC-nets may be practical
	for realistic applications, such as anonymous microblogging.
\end{compactitem}

Section~\ref{sec:bg}
introduces DC-nets and the disruption problem.
Section~\ref{sec:arch} outlines \app's architecture
and adversary model, and
Sections~\ref{sec:proto} and~\ref{sec:crypto} describe 
its messaging protocol and cryptographic schemes.
Section~\ref{sec:eval} presents
application scenarios and evaluation results,
Section~\ref{sec:rel} describes related work,
and Section~\ref{sec:concl} concludes.

%% file: bg.tex
\section{Background and Motivation}
\label{sec:bg}

This section first introduces the basic DC-nets concept
and known generalizations,
then motivates the need for proactive accountability.

\subsection{Anonymity with Strong Adversaries}
\xxx{Does this section belong here?}
To make the need for traffic-analysis-resistant anonymity systems more 
concrete, consider a political journalist 
who obtains some important secret government 
documents (e.g., the {\em Pentagon Papers}) 
from a confidential source.
If the journalist publishes these documents under her own name, 
the journalist might risk prosecution or interrogation,
and she might be pressured to reveal the source of the documents.

To reduce such risks,
a number of political journalists
could form a \app communication group.
Any participating journalist may then {\em anonymously broadcast} the documents
to the entire group of journalists, 
such that no member of the group can determine which journalist sent the documents.
With \app, even if a government agency plants agents within the
group of journalists and observes {\em all network traffic} during 
a protocol run, the agency remains unable to learn
the source of the leak.

Existing systems such as Tor,
which are practical and scalable but vulnerable to known traffic analysis
attacks~\cite{dingledine04tor,moeller00mixmaster,danezis03mixminion},
cannot guarantee security in this context.
For example,
if a US journalist posts a leak to a US website,
via a Tor connection whose entry and exit relays are in Europe,
then an eavesdropper capable of
monitoring transatlantic links~\cite{macaskill13gchq}
can de-anonymize the user
via traffic analysis~\cite{dingledine04tor,murdoch07sampled}.
Prior anonymity systems attempting to offer resistance to traffic analysis,
discussed in Section~\ref{sec:rel}, suffer from poor performance
or vulnerability to active denial-of-service attacks.

\subsection{DC-nets Overview}
\label{sec:bg:overview}

\begin{figure}[t]
\centering
\includegraphics[width=0.45\textwidth]{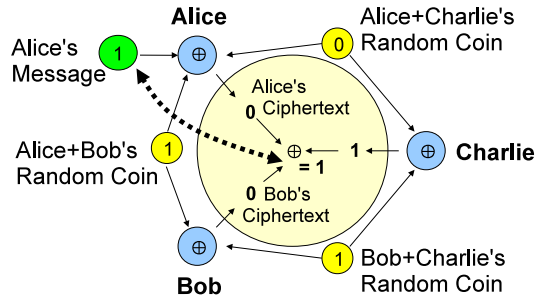}
\caption{The basic DC-nets algorithm}
\label{fig:dcnets}
\end{figure}

DC-nets~\cite{chaum88dining}
provide anonymous broadcast within a 
{\em group} of participants,
who communicate lock-step in a series of {\em rounds}.
In a given round,
each group member contributes an equal length ciphertext
that, when combined with all other members' ciphertexts,
reveals one or more cleartext messages.
All group members know that each message was sent
by {\em some} group member---%
but do not know {\em which} member sent each message.

In its simplest form,
illustrated in Figure~\ref{fig:dcnets},
we assume one group member wishes to broadcast a 1-bit message anonymously.
To do so, every pair of members flips a coin,
secretly agreeing on the random outcome of that coin flip.
An $N$-member group thus flips $N(N-1)/2$ coins in total,
of which each member observes the outcome of $N-1$ coins.
Each member then XORs together the values of the $N-1$ coins she observes,
additionally the member who wishes to broadcast the 1-bit message
XORs in the value of that message,
to produce that member's DC-nets {\em ciphertext}.
Each group member then broadcasts her 1-bit ciphertext to the other members.
Finally, each member collects and XORs all $N$ members' ciphertexts together.
Since the value of each shared coin is XORed into exactly two members' ciphertexts,
all the coins cancel out, leaving only the anonymous message,
while provably revealing no information
about which group member sent the message.

\subsection{Practical Generalizations}

As a standard generalization of DC-nets to communicate $L$-bit messages,
all members in principle simply run $L$ instances of the protocol in parallel.
Each pair of members flips and agrees upon $L$ shared coins,
and each member XORs together the $L$-bit strings she observes
with her optional $L$-bit anonymous message to produce $L$-bit ciphertexts,
which XOR together to reveal the $L$-bit message.
For efficiency,
in practice each pair of group members forms a cryptographic shared secret---%
via Diffie-Hellman key agreement, for example---%
then group members use a cryptographic pseudo-random number generator (PRNG)
to produce the $L$-bit strings.

As a complementary generalization,
we can use any finite alphabet or group in place of coins or bits,
as long as we have:
(a) a suitable combining operator analogous to XOR,
(b) a way to encode messages in the chosen alphabet,
and
(c) a way to generate complementary pairs of one-time pads in the alphabet
that cancel under the chosen combining operator.
For example, the alphabet might be 8-bit bytes,
the combining operator might be addition modulo 256,
and from each pairwise shared secret,
one member of the pair generates bytes $B_1,\dots,B_k$ from a PRNG,
while the other member generates corresponding two's complement bytes
$-B_1,\dots,-B_k$.

\subsection{Disruption and Verifiable DC-nets}

A key weakness of DC-nets is that a
single malicious insider can easily block 
all communication.
An attacker who transmits arbitrary bits---instead of the 
XORed ciphertext that the protocol prescribes---%
unilaterally and {\em anonymously} jams
all DC-net communication. 

\label{sec:bg-benefits}

In many online venues such as blogs, chat rooms, and social networks,
some users may have legitimate needs for strong anonymity---%
protest organizers residing in an authoritarian state, for example---%
while other antagonistic users (e.g., secret police infiltrators)
may attempt to block communication
if they cannot de-anonymize ``unapproved'' senders.
Even in a system like Dissent
that can {\em eventually} trace and exclude disruptors,
an adversary with multiple colluding dishonest group members
may still be able to slow or halt communication for long enough
to ruin the service's usability for honest participants.
Further, if the group's membership is open enough to allow
new disruptive members to join more quickly than the tracing process operates,
then these infiltrators may be able to shut down communication permanently.

Verifiable DC-nets~\cite{golle04dining}
leverage algebraic groups, such as elliptic curve groups, as the DC-nets alphabet.
Using such groups allows for disruption resistance,
by enabling members to {\em prove} the correctness
of their ciphertexts' construction
without compromising the secrecy of the shared pseudo-random seeds.
Using a hybrid approach that combines a traditional DC-net with a verifiable
DC-net, \app can achieve the messaging latency of a traditional XOR-based
DC-net while providing the strong disruption-resistance of verifiable DC-nets.

%% file: arch.tex
\section{\App Architecture Overview}
\label{sec:arch}

\begin{figure*}
\centering
\begin{tabular}{ccc}
\includegraphics[height=0.85in]{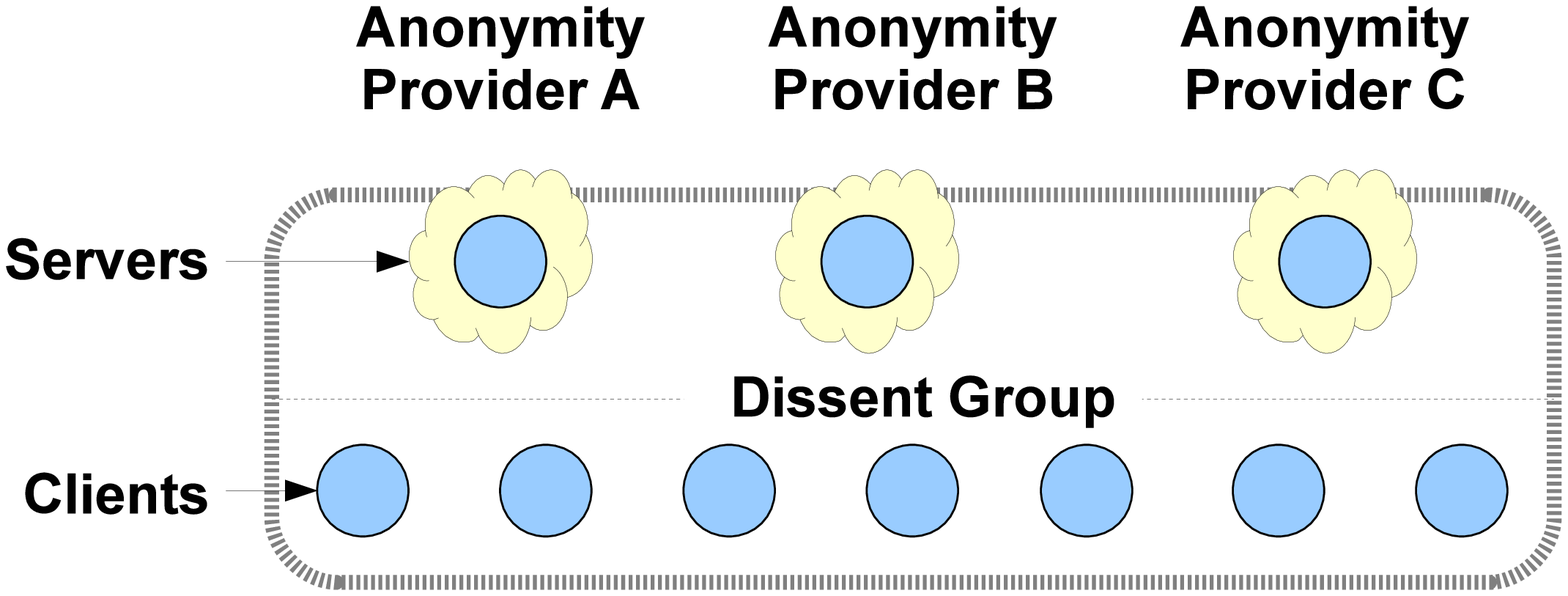}	&
\includegraphics[height=0.85in]{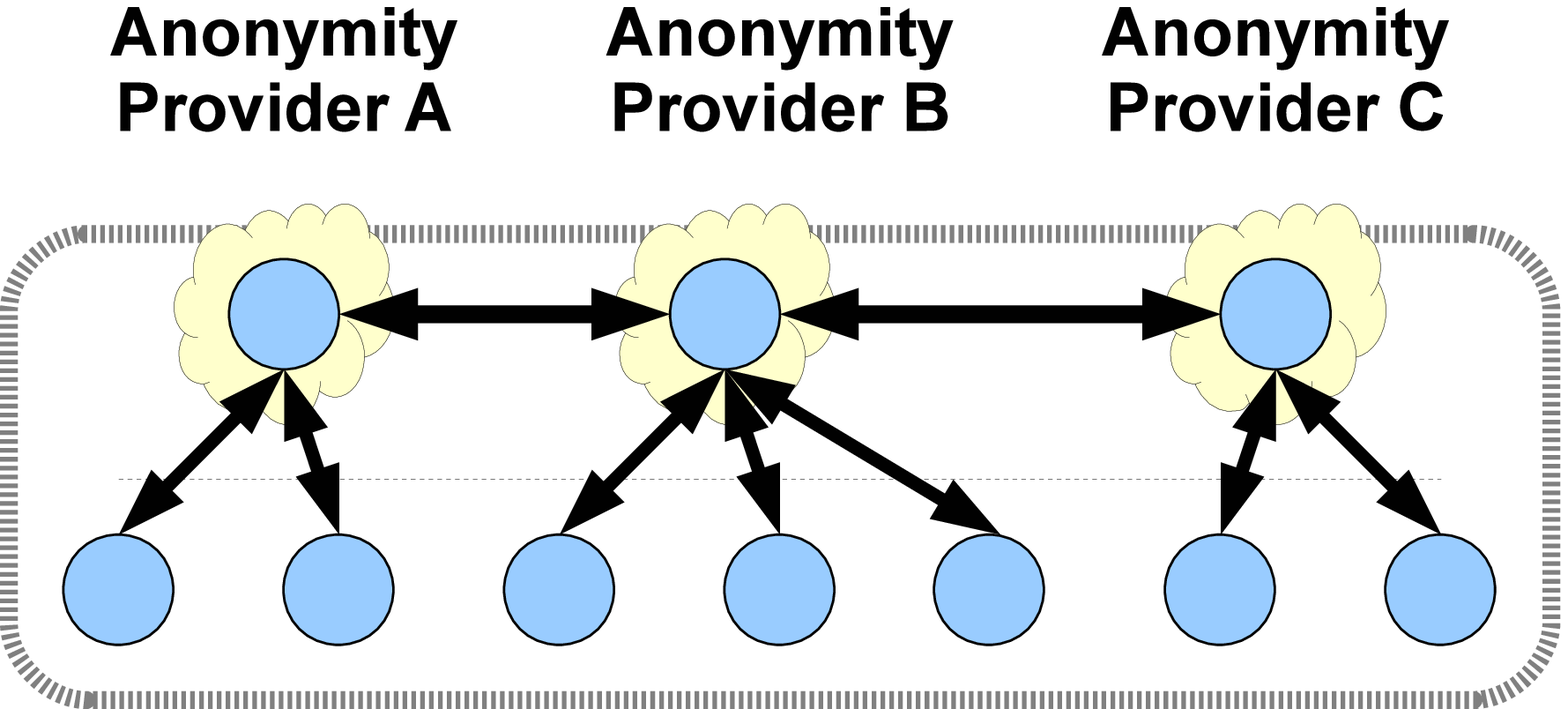}	&
\includegraphics[height=0.85in]{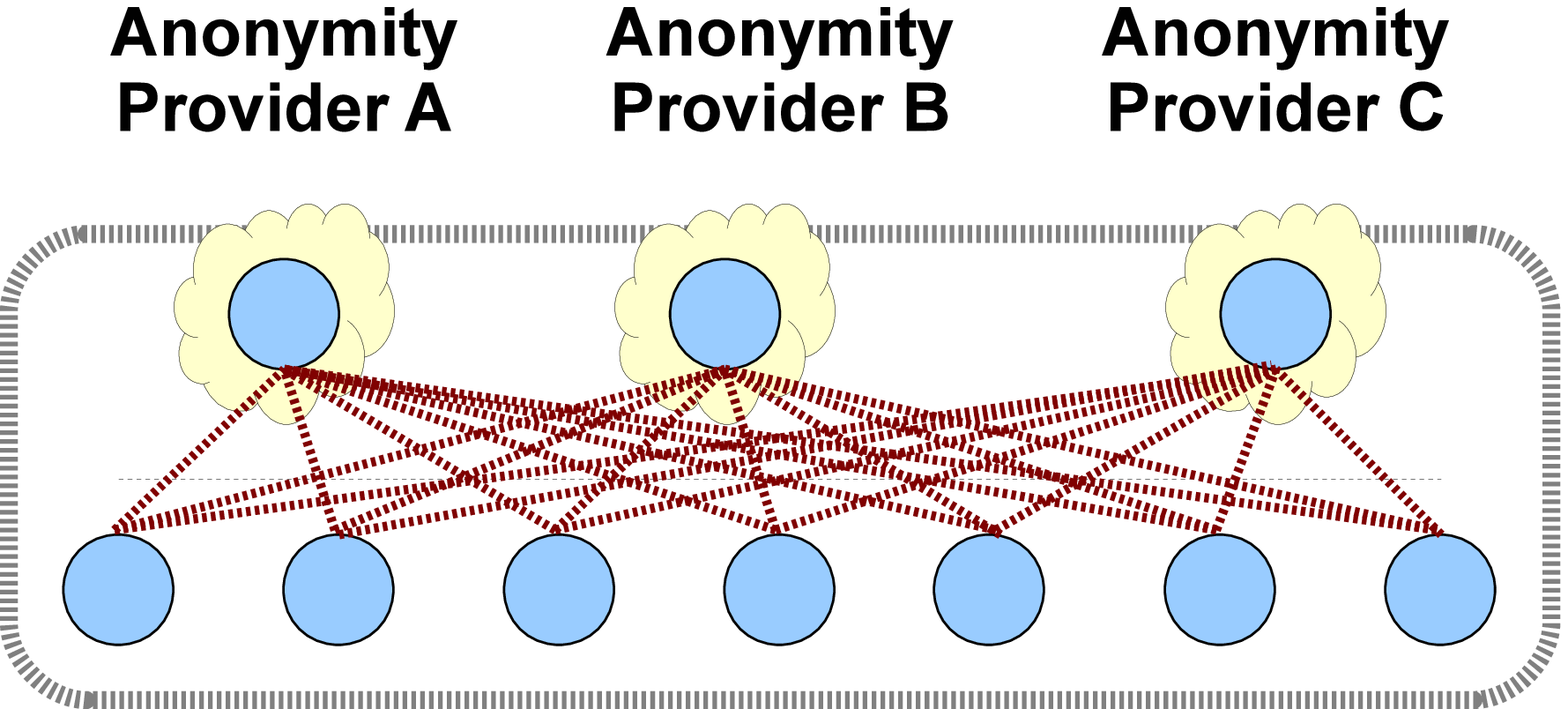}	\\
(a) Multi-provider cloud model~\cite{wolinsky12dissent}		&
(b) Communication topology					&
(c) DC-nets secret sharing 					\\
\end{tabular}
\caption{\App/Dissent deployment model,
	physical communication topology,
	and DC-nets secret sharing}
\label{fig:anytrust}
\end{figure*}

In this section, we describe the individual components
of \app and how they combine to form the overall
anonymous communication system.

\subsection{Deployment and Adversary Model} 
\label{sec:arch-cs}

\App builds on Dissent~\cite{wolinsky12scalable, wolinsky12dissent}
and uses the {\em multi-provider cloud} model
illustrated in Figure~\ref{fig:anytrust}~(a)
to achieve scalability and resilience to ordinary node and link failures.
In this model, a communication group consists of mostly unreliable {\em clients},
and a few {\em servers} we assume to be
highly available and well-provisioned.
Servers in a group should be administered independently---%
each furnished by a different {\em anonymity provider}, for example---%
to limit risk of all servers being compromised
and colluding against the clients.
The servers may be geographically or topologically close, however---%
possibly even hosted in the same data center,
in locked cages physically and administratively accessible
only to separate, independent authorities.

Clients directly communicate, at a minimum, with a single upstream server,
while each server communicates with all other servers.
This topology, shown in Figure~\ref{fig:anytrust}~(b),
reduces the communication and computation burden on the clients,
and enables the system to make progress regardless of client churn.
In particular, clients need not know which other clients are online
at the time they submit their DC-net ciphertexts to their upstream server;
clients only assume that all {\em servers} are online.

To ensure anonymity,
clients need {\em not} assume that any particular server
is trustworthy---a client need not even trust its
immediately upstream server.
Instead, {\bf clients trust only that there exists
{\em at least one} one honest server},
an assumption previously dubbed
{\em anytrust}~\cite{wolinsky12scalable,wolinsky12dissent},
as a trust analog to anycast communication.

\app, like Dissent,
achieves security under the anytrust assumption
through the DC-nets key-sharing model
shown in Figure~\ref{fig:anytrust}~(c).
Each client shares a secret with {\em every} server,
rendering client ciphertexts indecipherable
without the cooperation of {\em all} servers,
and hence protecting a client's anonymity even if
its immediately upstream server is malicious.
Each client ultimately obtains an anonymity set
consisting of the set of all honest clients,
provided that the anytrust assumption holds,
and provided the message contents themselves 
do not in some way reveal the sender's identity.

A malicious server might refuse to service honest clients,
but such refusal does not compromise clients' anonymity---%
victims can simply switch to a different server.
Although not yet supported in our \app prototype,
Section~\ref{sec:proto:future} discusses how
one might use threshold secret sharing
to tolerate server failures,
at the cost of requiring that we assume
multiple servers are honest.

\subsection{Security Goals}

\app's goal is to offer anonymity and disruption resistance
in the face of a strong adversary
who can potentially monitor all network links,
modify packets as they traverse the network, and 
compromise a potentially large fraction of a group's participating members.
We say that a participant is {\em honest} if it
follows the protocol exactly and does not collude
with or leak secret information to other nodes.
A participant is {\em dishonest} otherwise.
Dishonest nodes can exhibit {\em Byzantine behavior}---%
they can be arbitrarily incorrect and
can even just ``go silent.'' 

The system is designed to provide anonymity
among the set of {\em honest} participants,
who remain online and uncompromised throughout an interaction period,
and who do not compromise their identity
via the content of the messages they send.
We define this set of honest and online participants
as the {\em anonymity set} for a protocol run.
If a group contains many colluding dishonest participants,
\app can anonymize the honest participants 
only among the remaining subset of {\em honest} members:
in the worst case of a group containing only one honest member, for example,
\app operates but can offer that member no meaningful anonymity.

Similarly, \app does not prevent
long-term intersection attacks~\cite{kedogan02limits}
against otherwise-honest participants who repeatedly come and go
during an interaction period,
leaking information to an adversary who can correlate
online status with linkable anonymous posts.
Reasoning about anonymity sets generally requires
making inherently untestable assumptions about {\em how many} group members
may be dishonest or unreliable,
but \app at least does not assume that the honest participants know {\em which}
other participants are honest and reliable.

Finally, \app's disruption-resistant design
addresses {\em internal} disruption attacks
by misbehaving anonymous participants,
a problem specific to anonymous 
communication tools and particularly DC-nets.
Like any distributed system, \app may be vulnerable to
more general network-level Denial-of-Service (DoS) attacks as well,
particularly against the servers that are critical
to the system's availability and performance.
Such attacks are important in practice,
but not specific to anonymous communication systems.
This paper thus does not address general DoS attacks
since well-known defenses apply,
such as server provisioning, selective traffic blocking,
and proof-of-life or proof-of-work challenges.

%% file: proto.tex
\section{Protocol Design}
\label{sec:proto}

\App consists of
two major components: the messaging protocol, and
the cryptographic primitive
clients and servers use to construct their ciphertexts.
This section describes the \app messaging protocols,
and the following section 
describes the cryptographic constructions.

\subsection{Core \App Protocol}
\label{sec:proto-basic}

\begin{figure}[t]
\begin{boxedminipage}{\linewidth}

\newcounter{phaseCounter}
\setcounter{phaseCounter}{0}
\newcommand{\phase}[1]{\item \refstepcounter{phaseCounter}%
  \arabic{phaseCounter}. \textbf{#1}.}

\begin{trivlist}

  \phase{Client Ciphertext Generation}
    \label{proto:client-ciphertext-gen}
    Each client $i$ generates a client ciphertext,
    and submits this ciphertext to client $i$'s upstream server.
    If client $i$ is the anonymous owner of the current slot,
    the client computes and submits a slot owner ciphertext
    using her pseudonym secret key and her plaintext message $m$.

  \phase{Client Set Sharing}
    \label{proto:server-ciphertext-share}
    After receiving valid client ciphertexts
    from its currently connected downstream clients,
    each server $j$ broadcasts to all servers
    its set $\textbf{C}_j$ of collected client ciphertexts.

  \phase{Server Ciphertext Generation}
    \label{proto:server-ciphertext-gen}
    After receiving client ciphertext sets from all servers,
    each server $j$ computes $\textbf{C} = \bigcup_k \textbf{C}_k$,
    the set of client ciphertexts collected by {\em all} servers. 
    Server $j$ then uses $\textbf{C}$
    to compute a server ciphertext corresponding to the set
    of submitted client ciphertexts.
    Server $j$ broadcasts this server ciphertext to all other servers.

  \phase{Plaintext Reveal}
    \label{proto:server-plaintext-reveal}
    After receiving a server ciphertext from every other server,
    each server $j$ combines the $|\textbf{C}|$ client ciphertexts
    and $M$ server ciphertexts to reveal the plaintext message $m$.
    Server $j$ signs $m$ and broadcasts its 
    signature $\sigma_j$ to all servers.

  \phase{Plaintext Sharing}
    \label{proto:server-plaintext-share}
    After receiving valid signatures from all servers,
    server $j$ sends the plaintext message $m$ and the $M$
    signatures $\sigma_1, \dots, \sigma_M$ (one from each server)
    to its downstream clients. 

  \phase{Client Verification}
    \label{proto:client-ver}
    Upon receiving the plaintext $m$ and 
    $M$ valid signatures from its upstream server,
    client $i$ accepts the plaintext message
    and considers the messaging round to have completed successfully.
\end{trivlist}

{\em All messages sent over the network include 
a session nonce and are signed with the sender's long-term 
well-known (non-anonymous) signing key.}
\end{boxedminipage}

\caption{Core \App messaging protocol}
\label{fig:protocol}
\end{figure}

Figure~\ref{fig:protocol} summarizes the
steps comprising a normal-case run of the \app protocol.
This protocol represents
a direct adaptation of the DC-nets scheme (Section~\ref{sec:bg:overview})
to the two-level communication topology
illustrated in Figure~\ref{fig:anytrust}~(b),
and to the client/server secret-sharing graph
in Figure~\ref{fig:anytrust}~(c).
As in Dissent, \app's anonymity guarantee
relies on Chaum's original security analysis~\cite{chaum88dining},
in which an honest node's anonymity set consists of
the set of honest nodes that remain connected in the secret-sharing graph
after removing links to dishonest nodes.
Since each client shares a secret with every server,
and we assume that there exists at least one honest server,
this honest server forms a ``hub'' connecting all honest nodes.
This anonymity property holds regardless of physical communication topology,
which is why the clients need not trust their immediately upstream server.

The ciphertext- and proof-generation processes assume
that communication in the DC-net is broken up into 
{\em time slots} (akin to TDMA), such that only one client---the slot's
{\em owner}---is allowed to send an anonymous message in each time slot.
The owner of a slot is the client who holds the 
private key corresponding to a {\em pseudonym public key}
assigned to the slot.
To maintain the slot owner's anonymity,
{\em no one must know} which client owns which transmission slot.
Section~\ref{sec:proto-many} below describes the
assignment of pseudonym keys to transmission slots.

\begin{figure}
  \centering
  \includegraphics[width=0.45\textwidth]{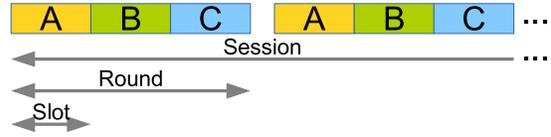}
  \caption{Example DC-net transmission schedule, where
    anonymous authors A, B, and C transmit in each round.}
  \label{fig:fig-slot}
\end{figure}

Figure~\ref{fig:fig-slot} shows an example DC-net transmission schedule
with three slots, owned by pseudonyms A, B, and C.
Each slot owner can transmit one message per {\em messaging round},
and the slot ordering in the 
schedule remains the same for the duration of the \app{} {\em session}.

\subsection{Verifiable Ciphertexts in \app}

While \app's anonymity derives from the same principles as Dissent's,
the key difference is in the ``alphabet''
with which \app generates DC-net ciphertexts,
and in the way \app combines and checks those ciphertexts.
Dissent uses a symmetric-key cryptographic
pseudo-random number generators (PRNG) to generate shared secrets,
and uses bitwise XOR to combine them and later to reveal the plaintext message.
While efficient, the symmetric-key approach offers no way to check that
any node's ciphertext was generated correctly
until the final cleartext messages are revealed.
If any node corrupts a protocol round by sending an incorrect ciphertext,
Dissent can eventually identify that node only via
a complex retroactive {\em blame} procedure.

\App, in contrast, divides messages into chunks small enough to be
encoded into elements of algebraic groups,
such as Schnorr~\cite{schnorr91efficient} or elliptic curve groups,
to which known proof-of-knowledge techniques apply.
Section~\ref{sec:crypto} later outlines three particular
ciphertext generation schemes that \app implements,
although \app's architecture and protocol design
is agnostic to the specific scheme.
These schemes may be considered analogous
to ``pluggable'' ciphersuites in SSL/TLS.

Thus, any \app ciphertext is generated {\em on behalf}
of the holder of some particular pseudonym keypair.
While the details of a ciphertext correctness proof
depend on the particular scheme,
all such proofs have the general form of an ``either/or'' knowledge proof,
of the type systematized by Camenisch and Stadler~\cite{camenisch97proof}.
In particular, a ciphertext correctness proof attests that either:
\begin{compactitem}
\item	the ciphertext is an encryption of {\em any} message,
	and the producer of the ciphertext holds
	the {\em private} part of the pseudonym key for this slot, OR
\item	the ciphertext is an encryption of a {\em null} cover message,
	which, when combined with other cover ciphertexts
	and exactly one actual encrypted message ciphertext,
	will combine to reveal the encrypted message.
\end{compactitem}

Only the pseudonym key owner can produce a correctness proof
for an arbitrary message following the first alternative above,
while any node can generate an ``honest'' cover ciphertext---%
and the proof by construction reveals no information
about {\em which} alternative the proof generator followed.
We leave discussion of further details of this process
to Section~\ref{sec:crypto},
but merely note that such ``either/or'' proofs are 
pervasive and well-understood in the cryptographic theory community.

In \app,
each client computes and attaches a cryptographic correctness proof
to each ciphertext it sends to its upstream server,
and each server in turn attaches a correctness proof
to the server-side ciphertext it generates in Phase~3 of each round
(Figure~\ref{fig:protocol}).
In principle, therefore, each server can immediately verify
the correctness of any client's or other server's ciphertext it receives,
{\em before} ``accepting'' it and combining it 
with the other ciphertexts for that protocol round.
As in Dissent, \app achieves resilience to client churn 
by (optionally) requiring clients to submit
their ciphertexts before a certain ``deadline'' in each messaging round.
We describe this technique in Section~\ref{sec:proto-silent}.

While \app nodes can {\em in principle} verify the correctness of 
any received ciphertext immediately,
actually doing so has performance costs.
These costs lead to design tradeoffs
between ``eager'' and ``lazy'' verification,
both of which we implement and evaluate later in Section~\ref{sec:eval}.
Lazy verification enables the critical servers
to avoid significant computation costs during rounds that are not disrupted,
at the expense of making a round's output unusable if it {\em is} disrupted.
Even if a lazily-verified round is disrupted, however,
the fact that \app nodes always generate and transmit 
signed ciphertext correctness proofs
greatly simplifies and shortens
the retroactive blame process with respect to Dissent.

\app currently treats {\em server-side} failures of all types,
including invalid server ciphertexts,
as ``major events'' requiring administrative action,
and simply halts the protocol with an alert until such action is taken.
Section~\ref{sec:bft} later discusses approaches
to make \app resilient against server-side failures,
but we leave implementing and evaluating such mechanisms to future work.
Such server-side failures affect only availability, however;
anonymity remains protected as long as at least one 
(not necessarily online) server remains uncompromised.

\subsection{Scheduling Pseudonym Keys}
\label{sec:proto-many}

To assign ownership of transmission slots to clients
in such a way that {\em no one} knows which client owns which slot,
\app applies an architectural idea from Dissent~\cite{wolinsky12dissent}.
At the start of a \app session, each of the $N$ clients secretly submits a
slot owner pseudonym key to a verifiable 
shuffle protocol~\cite{neff01verifiable} run by the servers.
The public output of the shuffle is the $N$ pseudonym
keys in permuted order---such that {\em no one} 
knows which node submitted which pseudonym key
other than their own.
\app participants then use each of 
these $N$ pseudonym keys to initialize
$N$ concurrent instances of the core \app DC-net
with each instance becoming a slot in a verifiable DC-net
{\em transmission schedule}.

\paragraph{Scheduling Policy}
Not every client will necessarily 
want to transmit an anonymous message
in every messaging round.
In addition, clients
may want to transmit messages of different lengths.
To make \app more efficient in these cases,
\app allows clients to request a change
in the length of their messaging slot (e.g.,
so that a client can send a long message in a single
messaging round) and to 
temporarily ``close'' their transmission slot
(if a client does not expect to send a message 
for several rounds).
Clients issue these requests
by prepending a few bits of control data
to the anonymous message they send in their
transmission slot.

\subsection{Hybrid XOR/Verifiable DC-Nets}
\label{sec:proto-hybrid}

While the verifiable DC-net design above
may be needed in scenarios in which disruptions are frequent,
the public-key cryptography involved imposes a much higher computational
cost than traditional XOR-based DC-nets.
To offer better performance
in groups with fewer or less frequent disruptions,
\app has a ``hybrid'' mode of operation 
that uses the fast XOR-based DC-net when there are
no active disruptors in the group, and
reverts to a verifiable DC-net in the presence
of an active disruptor.
This hybrid \app DC-net marries the relatively 
low computational cost of the XOR-based DC-net with 
the low accountability cost of the verifiable DC-net.

To set up a hybrid DC-net session, all protocol
participants first participate in a pseudonym signing
key shuffle, as described above in Section~\ref{sec:proto-many}.
At the conclusion of the shuffle, all nodes initialize
{\em two} DC-net slots for {\em each} of the
$N$ clients: one traditional Dissent-style DC-net,
and one verifiable \app DC-net.

When the group is not being disrupted,
clients transmit in their Dissent DC-net slot, 
allowing nodes to take advantage of the speed
of Dissent's XOR-based DC-net. 
When nodes detect the corruption of a message in the Dissent
DC-net, the client whose message was corrupted reverts
to transmitting in its {\em verifiable} DC-net slot.
This client can use the verifiable slot to 
transmit anonymously the ``accusation'' necessary to 
identify the disruptor in the 
Dissent accusation process~\cite[Section 3.9]{wolinsky12dissent}.
The \app DC-net replaces
the expensive verifiable shuffle necessary for nodes
to exchange the accusation information in Dissent.
In this way, \app offers the normal-case efficiency of XOR-based DC-nets
while greatly reducing the cost of tracing and excluding disruptors.

\subsection{Client Churn}
\label{sec:proto-silent}

In realistic deployments
clients may go offline at any time,
and this problem becomes severe in large groups
of unreliable clients exhibiting constant churn.
To prevent slow or unresponsive clients from disrupting communication,
the servers need not wait in Phase~\ref{proto:server-ciphertext-share}
for all downstream clients to submit ciphertexts.
Instead, servers can wait
for a fixed threshold of $t \leq N$ clients
to submit ciphertexts, or
for some fixed time interval $\tau$.
Servers might also use some more complicated 
{\em window closure policy}, as in Dissent~\cite{wolinsky12dissent}:
e.g., wait for a threshold of clients and then
an additional time period before proceeding.
The participants must agree on a window closure policy
before the protocol run begins.

There is an inherent tradeoff between anonymity and
the system's ability to cope with unresponsive clients.
If the servers close the ciphertext submission window too aggressively,
honest but slow clients
might be unable to submit their ciphertexts in time,
reducing the anonymity of clients who do manage to submit messages.
In contrast, if the servers wait until 
every client has submitted a ciphertext, a
single faulty client could prevent protocol
progress indefinitely.
Optimal policy choices depend on the
security requirements of the application at hand.

\subsection{Limitations and Future Enhancements}
\label{sec:proto:future}
This section outlines some of \app's current limitations, deployment issues,
and areas for future work.

\paragraph{Group Evolution}
\App's architecture assumes that, at the start of the protocol, group members
agree to a ``roster'' of protocol participants---%
essentially a list of public keys defining the group's membership.
The current prototype simplistically assumes that
this group roster is a static list,
and that the session nonce is a hash of a file
containing this roster and other group policy information.
This design trivially ensures that all nodes participating in a given group
(uniquely identified by its session nonce) agree upon the same roster and policy.
Changing the group roster or policy in the current prototype
requires forming a new group, but 
we are exploring approaches to group management
which would allow for on-the-fly membership changes.
For now, we simply note that \app's security 
properties critically depend on group membership policy decisions,
which affect how quickly adversarial participants can infiltrate a group.
We consider group management policy to be orthogonal
to this paper's communication mechanisms.

\paragraph{Sybil Attacks}
If it is too easy to join a group, dishonest participants
might flood the group with Sybil identities~\cite{douceur02sybil},
giving an anonymous slot owner the impression 
that she has more anonymity 
than she actually does.
The current ``static group'' design 
shifts the Sybil attack prevention problem
to whomever formulates the group roster.
Dynamic group management schemes could leverage
existing Sybil prevention
techniques~\cite{tran09sybil,yu08sybillimit,yu06sybilguard},
but we do not consider such approaches herein.

\paragraph{Membership Concealment}
\app considers the group roster,
containing the public keys of all participants,
to be public information: concealing 
participation in the protocol is an orthogonal security goal
that \app currently does not address.
We are exploring the use of anonymous 
authentication techniques~\cite{fujisaki07traceable,liu04linkable,rivest01leak}
to enable \app clients to ``sign on'' and prove membership in the group
without revealing to the \app servers (or to the adversary)
{\em which} specific group members are online at any given time.
Further, we expect that \app's design could be composed with other techniques
to achieve membership 
concealment~\cite{moghaddam12skypemorph,vasserman09membership},
but we leave such enhancements to future work.

\paragraph{Unresponsive Servers}
\App currently assumes that the servers supporting a group
are well-provisioned and highly reliable,
and the system simply ceases communication progress
in the face of any server's failure.
Any fault-masking mechanism would be problematic, in fact,
in the face of \app's assumption that only one server might be honest:
if that one honest server goes offline and the protocol continues without it,
then the remaining dishonest servers could collude against all honest users.

If we assume that there are
$h > 1$ honest servers, however, a 
currently unimplemented variation of \app could
allow progress if as many as $h-1$ servers are unresponsive.
Before the protocol run, every server
uses a {\em publicly verifiable 
secret sharing scheme}~\cite{schoenmakers99simple},
to distribute shares of its per-session secret key.
The scheme is configured such that
any quorum of $M - h + 1$ shares is
sufficient to reconstruct the secret.
Thus, at least one honest server must remain online
and contribute a share for a secret to be reconstructed.
(Golle and Juels~\cite{golle04dining} also use
a secret-sharing scheme, but they
rely on a trusted third-party to 
generate and distribute the shares.)

If a server becomes unresponsive,
the remaining online servers can broadcast their shares
of the unresponsive server's secret key.
Once a quorum of servers broadcasts these shares, the remaining
online servers will be able to reconstruct the unresponsive
server's private key.
Thereafter, each server can simulate the
unresponsive server's messages for
the rest of the protocol session.

\paragraph{Blame Recovery}
\label{sec:bft}

The current \app prototype can detect server misbehavior, but
it does not yet have a mechanism by which the remaining servers can 
collectively form a new group ``roster'' with the misbehaving nodes removed.
We expect off-the-shelf Byzantine Fault Tolerance
algorithms~\cite{castro99practical} 
to be applicable to this {\em group evolution} problem.
Using BFT to achieve agreement, however,
requires a stronger security assumption:
in a group with $f$ dishonest servers, there
must be at least $3f+1$ total servers.
We sketch a possible BFT-based group evolution approach here.

The BFT cluster's shared state in this case
is the group ``roster,'' containing the session nonce and 
a list of all active \app 
clients and servers, identified by their public keys.

The two operations in this BFT system are:
\begin{compactitem}
  \item \texttt{EVOLVE\_GROUP(nonce, node\_index, proof)},
    a request to remove a dishonest node
    (identified by \texttt{node\_index}) from the group roster.
    BFT servers remove the dishonest node from the group 
    if the proof is valid, yielding the new group roster.
  \item \texttt{GET\_GROUP()},
    which returns current the group roster.
\end{compactitem}
If, at some point during the \app session, a \app
node concludes that the protocol has failed due
to the dishonesty of node $d$,
this honest node makes an $\texttt{EVOLVE\_GROUP}$ 
request to the BFT cluster and waits for a response.
The honest BFT servers will agree on a new group
roster with the dishonest node $d$ removed
and the \app servers will
begin a new instance of the \app protocol with the
new group roster.
Clients use $\texttt{GET\_GROUP}$ to learn the
new group roster.

%% file: crypto.tex
\section{Verifiable DC-net Constructions}
\label{sec:crypto}
The \app architecture relies on
a verifiable DC-net primitive that has many possible implementations.
In this section, we first describe the general interface that
each of the cryptographic constructions must implement---%
which could be described as a ``\app ciphersuite API''---%
then we describe the three specific experimental schemes
that \app currently implements.

All three schemes are founded on standard, well-understood
cryptographic techniques
that have been formally developed and rigorously analyzed in prior work.
As with most practical, complex distributed systems with many components,
however,
we cannot realistically offer
a rigorous proof that these cryptographic tools ``fit together''
correctly to form a perfectly secure overall system.
(This is true even of SSL/TLS and its ciphersuites,
which have received far more cryptographic scrutiny than \app
but in which flaws are still found regularly.)
We also make no claim that any of the current schemes
are ``the right'' ones or achieve any particular ideal;
we merely offer them as contrasting points in a large design space.
To lend some informal credibility to their security,
we note that our pairing-based scheme is closely modeled on
the verifiable DC-nets scheme that Golle and Juels
already developed formally~\cite{golle04dining},
and \abbr{the extended version of this paper~\cite{corrigangibbs13proactively-tr}}{Appendix~\ref{app:proof}} 
sketches a security argument
for the third and most computationally efficient scheme.

\subsection{Verifiable DC-net Primitive API}
\label{sec:crypto-prim}

The core cryptographic primitive 
consists of a set of six methods. 
Each of these six methods takes a list of protocol session parameters
(e.g., group roster, session nonce, slot owner's
public key) as an implicit argument:
\begin{compactitem}
  \item \textit{Cover Create:} 
    Given a client session secret key, return 
    a valid client ciphertext representing ``cover traffic,''
    which do not contain actual messages.

  \item \textit{Owner Create:} 
    Given a client session secret key, 
    the slot owner's pseudonym secret key,
    and a plaintext message $m$
    to be transmitted anonymously, 
    return a valid {\em owner} ciphertext that encodes message $m$.

  \item \textit{Client Verify:} 
    Given a client public key and a client
    ciphertext, return a boolean flag
    indicating whether the client ciphertext is
    valid.

  \item \textit{Server Create}: 
    Given a server private key and
    a set of client ciphertexts,
    return a valid server ciphertext.

  \item \textit{Server Verify:}
    Given a server public key, a set of valid client
    ciphertexts, and a server ciphertext, return a
    flag indicating whether the server
    ciphertext is valid.

  \item \textit{Reveal:}
    Combine $N$ client ciphertexts
    and $M$ server ciphertexts,
    returning the plaintext message $m$.
\end{compactitem}
However these methods are implemented, 
they must obey the following 
security properties, stated informally: 
\begin{compactitem}
  \item \textbf{Completeness:}
        An honest verifier always accepts a ciphertext
        generated by an honest client or server.

  \item \textbf{Soundness:}
        With overwhelming probability an honest verifier rejects
        an incorrect ciphertext, such as an owner ciphertext
        formed without knowledge of the owner's pseudonym secret key. 

  \item \textbf{Zero-knowledge:}
        A verifier learns nothing about a ciphertext besides
        the fact that it is correctly formed.

  \item \textbf{Integrity:}
        Combining $N$ valid client ciphertexts, including one ciphertext
        from the anonymous slot owner, and $M$ valid server ciphertexts,
        reveals the slot owner's plaintext message.

  \item \textbf{Anonymity:}
        A verifier cannot distinguish a client ciphertext from 
        the anonymous slot owner's ciphertext.
        \abbr{The extended version of this paper~\cite{corrigangibbs13proactively-tr}}{Appendix~\ref{app:proof}}
        offers a game-based definition of anonymity.
\end{compactitem}
In practice, each of our current implementations of this
verifiable DC-nets primitive achieve these security properties
in the random-oracle model~\cite{bellare93random} 
using non-interactive zero-knowledge proofs~\cite{groth04honest}.

\subsection{ElGamal-Style Construction}
\label{sec:crypto-elgamal}

The first scheme builds on
the ElGamal public-key cryptosystem~\cite{elgamal85public}.
In ElGamal, a public/private keypair has the form 
$\langle B, b \rangle = \langle g^b, b \rangle$,%
\footnote{
We do not require that a trusted third party generate
participants' keypairs, but we {\em do} require participants
to prove knowledge of their secret key at the start 
of a protocol session, for reasons described in
\abbr{the extended version of this paper~\cite{corrigangibbs13proactively-tr}}{Appendix~\ref{app:malkeys}}.
}
and plaintexts and ciphertexts are elements of an algebraic group $G$.%
\footnote{
Throughout,
unless otherwise noted,
group elements are members of a
finite cyclic group $G$ in which the Decision Diffie-Hellman (DDH)
problem~\cite{boneh98decision} is assumed computationally infeasible,
and $g$ is a public generator of $G$.
}
We refer to this as the ``ElGamal-style'' construction because
its use of an ephemeral public key and encryption by multiplication
structurally resembles the ElGamal cryptosystem.
Our construction does {\em not} exhibit the malleability of 
textbook ElGamal encryption, however, because a proof of knowledge
of the secret ephemeral public key is attached 
to every ciphertext element.

\paragraph{Client Ciphertext Construction}
Given a list of server public keys $\langle B_1, \dots, B_M \rangle$,
a client constructs a ciphertext by selecting an ephemeral public 
key $R_i = g^{r_i}$ at random and computing the ciphertext element:
\begin{align*}
  C_i = m\left(\Pi_{j=1}^M B_j\right)^{r_i}
\end{align*}
If the client is the slot owner, the client
sets $m$ to its secret message, otherwise the client sets $m=1$.

To satisfy the security properties described in 
Section~\ref{sec:crypto-prim}, the client must somehow
{\em prove} that the ciphertext tuple $\langle R_i, C_i \rangle$ 
was generated correctly.
We adopt the technique of Golle and Juels~\cite{golle04dining}
and use a non-interactive proof-of-knowledge
of discrete logarithms~\cite{camenisch97proof}
to prove that the ciphertext has the correct form.
If the slot owner's pseudonym public key is $Y$, 
the client's ephemeral public key is $R_i$, and
the client's ciphertext element is $C_i$, the client generates a
proof:
\begin{align*}
  \textsf{PoK}\{r_i, y:
    \left( 
        R_i = g^{r_i} \land C_i = (\Pi_{j=1}^M B_j )^{r_i}
    \right)
  \lor Y = g^y\}
\end{align*}
In words: the sender demonstrates that {\em either}
it knows the discrete logarithm $r_i$ of the ephemeral public key $R_i$,
and the ciphertext is the product of all server public keys raised
to the exponent $r_i$;
{\em or} the sender knows the slot owner's secret pseudonym key $y$,
in which case the slot owner can set $C_i$ to a value of her choosing.
\abbr{The extended version of this paper~\cite{corrigangibbs13proactively-tr}}{Appendix~\ref{app:zkp}}
details how to construct
and verify this type of non-interactive zero-knowledge proof.

Note that a dishonest slot owner can set $C_i$ to a maliciously
constructed value (e.g., $C_i = 1$). 
The only effect of such an ``attack'' is that the slot owner
compromises {\em her own} anonymity.
Since a dishonest slot owner can always compromise her own anonymity
(e.g., by publishing her secret keys), a dishonest slot owner achieves
nothing by setting $C_i$ maliciously.

The tuple $\langle R_i, C_i, \textsf{PoK} \rangle$ serves 
as the client's ciphertext. 
As explained in Section~\ref{sec:proto-basic}, 
all participants sign the messages they exchange
for accountability.

\paragraph{Server Ciphertext Construction}

Given a server public key $B_j = g^{b_j}$ and
a list of ephemeral client public keys $\langle R_1, \dots, R_N \rangle$,
server $j$ generates its server ciphertext as:
\begin{align*}
  S_j = \left(\Pi_{i=1}^N R_i\right)^{-b_j}
\end{align*}
The server proves the validity of its ciphertext by 
creating a non-interactive proof of knowledge that it knows
its secret private key $b_j$ and that its ciphertext element
$S_j$ is the product of the ephemeral client keys raised 
to the exponent $-b_j$:
\begin{align*}
  \textsf{PoK}\{b_j:
    B_j = g^{b_j} \land
	  S_j = (\Pi_{i=1}^N R_i)^{-b_j} 
  \}
\end{align*}

\paragraph{Message Reveal}
To reveal the plaintext message, a participant 
computes the product of $N$ client ciphertext elements and
$M$ server ciphertext elements: 
\begin{align*}
  m = \left(\Pi_{i=1}^N C_i \right)\left(\Pi_{j=1}^M S_j \right)
\end{align*}
Each factor $g^{r_i b_j}$, where $r_i$ is client $i$'s ephemeral 
secret key and $b_j$ is server $j$'s secret key,
is included exactly twice in the above product---%
once with a positive sign in the client ciphertexts and 
once with a negative sign in the server ciphertexts.
These values therefore cancel, leaving only the plaintext $m$.

\paragraph{Drawbacks}
Since the clients must use a new ephemeral public key 
for {\em each} ciphertext element, sending a plaintext message 
that is $L$ group elements in length requires each client to
generate and transmit $L$ ephemeral public keys.
The proof of knowledge for this construction
is $L + O(1)$ group elements long,
so a message of $L$ group elements expands
to $3L + O(1)$ elements.

\subsection{Pairing-Based Construction}
\label{sec:crypto-pairing}
A major drawback of the ElGamal construction is that,
due to the need for ephemeral keys,
every ciphertext is three times as long as 
the plaintext it encodes.
Golle and Juels~\cite{golle04dining} use bilinear maps
to eliminate the need for ephemeral keys.
Our pairing-based construction adopts 
elements of their technique, while avoiding
their reliance on a trusted third party,
a secret-sharing scheme, and a probabilistic
transmission scheduling algorithm.

A symmetric bilinear map $\hat{e}$ maps two elements of
a group $G_1$ into a target group $G_2$---%
$\hat{e}: G_1 \times G_1 \rightarrow G_2$.
A bilinear map has the property that:
$\hat{e}(aP, bQ) = \hat{e}(P, Q)^{ab}$.\footnote{
Since $G_1$ is usually an elliptic curve group,
the generator of $G_1$ is written as $P$ 
(an elliptic curve point) and the repeated group operation
is written as $aP$ instead of $g^a$.
We will use the latter notation for consistency
with the rest of this section.
} 
To be useful, the map must also be non-degenerate
(if $P$ is a generator of $G_1$, $\hat{e}(P,P)$ is a generator of $G_2$)
and efficiently computable~\cite{boneh01identity}.
We assume that the decision bilinear Diffie-Hellman
assumption~\cite{boneh04efficient}  holds in $G_1$.\footnote{
Note that the decision Diffie-Hellman problem is easy
in $G_1$, since given $g, g^a, g^b, g^c \in G_1$, a DDH tuple
will always satisfy $\hat{e}(g^a, g^b) = \hat{e}(g, g^c)$
if $c = ab \bmod q$.
}

Since pairing allows a {\em single} pair of public keys to generate
a {\em sequence} of shared secrets, clients need not generate ephemeral
public keys for each ciphertext element they send.
This optimization leads to shorter ciphertexts
and shorter correctness proofs.

\paragraph{Client Ciphertext Construction}
For a set of server public keys $\langle B_1, \dots, B_M \rangle$,
a public nonce $\tau \in G_1$ computed using a hash function, and
a client public key $A_i = g^{a_i}$,
a pairing-based client ciphertext has the form:
\begin{align*}
  C_i = m \, \hat{e}(\Pi_{j=1}^M B_j, \tau)^{a_i}
\end{align*}
As before, if the client is not the slot owner,
the client sets $m = 1$. 
Each client can produce a proof of the correctness of its
ciphertext by executing a proof of knowledge similar to 
one used in the ElGamal-style construction above:
\begin{align*}
  \textsf{PoK}\{a_i, y:
	(  A_i = g^{a_i} \land C_i = \hat{e}(\Pi_{j=1}^M B_j, \tau)^{a_i} )
  \lor Y = g^y\}
\end{align*}

While the ElGamal-style scheme requires
$3L + O(1)$ group elements to encode $L$ elements of plaintext,
a pairing-based ciphertext requires only $L + O(1)$ group elements to encode
an $L$-element plaintext.

\paragraph{Server Ciphertext Construction}
Using a server public key $B_j=g^{b_j}$, 
a public round nonce $\tau$,
and client public keys
$\langle A_1, \dots, A_N \rangle$,
a server ciphertext has the form:
\begin{align*}
  S_j = \hat{e}(\Pi_{i=1}^N A_i, \tau)^{-{b_j}}
\end{align*}
The server proof of correctness is then:
\begin{align*}
  \textsf{PoK}\{b_j: 
    B_j = g^{b_j}
    \land 
    S_j = \hat{e}(\Pi_{i=1}^N A_i, \tau)^{-b_j} \}
\end{align*}

\paragraph{Message Reveal}
To reveal the plaintext,
the servers take the product of all
client and server ciphertexts:
\begin{align*}
  m = (\Pi_{i=1}^N C_i )(\Pi_{j=1}^M S_j)
\end{align*}

\paragraph{Drawbacks}
The main downside of this construction is
the relatively high computational cost of the pairing operation.
Computing the pairing operation on two 
elements of $G_1$ can take an order of magnitude
longer than a normal elliptic curve point addition
in a group of similar security level,
as Section~\ref{sec:eval-microbench} below will show.

\subsection{Hashing-Generator Construction}
\label{sec:crypto-hashing}

Our hashing-generator construction pursues a
``best of both worlds'' combination of the ElGamal-style
and pairing-based constructions.
This construction has short ciphertexts, like the
pairing-based construction, but avoids the computational
cost of the pairing-based scheme by using
conventional integer or elliptic curve groups.
To achieve both of these desired properties, the hashing-generator
construction adds some protocol complexity, in the form
of a session set-up phase.

\paragraph{Set-up Phase}

In the set-up phase,
each client $i$ establishes a Diffie-Hellman
shared secret $r_{ij}$ with every server $j$
using their respective public keys $g^{a_i}$ and $g^{b_j}$
by computing $r_{ij} = \textsf{KDF}(g^{a_ib_j})$
using a key derivation function $\textsf{KDF}$.
Clients publish commitments to these shared secrets
$R_{ij} = \hat{g}^{r_{ij}}$ 
using another public generator $\hat{g}$.

The hashing-generator construction
requires a process by which participants compute
a sequence of generators $g_1, \dots, g_L$ of the group $G$,
such that no participant knows the discrete logarithm of any 
of these generators with respect to any other generator.
In other words, {\em no one} knows an $x$ such that
$g_i^x = g_j$, for any $i,j$ pair.
In practice, participants compute this sequence of
generators by hashing a series of strings,
(e.g., the round nonce concatenated with ``1'', ``2'', ``3'', \dots), 
to choose the set of generating group elements.

At the end of the set-up phase, every client $i$ can produce a {\em sequence}
of shared secrets with each server $j$ using their shared secret
$r_{ij}$ and the $L$ generators: $g_1^{r_{ij}}, \dots, g_L^{r_{ij}}$.
In the $\ell$th message exchange round, all participants 
use generator $g_\ell$ as their common generator.

\paragraph{Client Ciphertext Construction}
To use the hashing-generator scheme to create a ciphertext,
the client uses its shared secrets $r_{i1}, \dots, r_{iM}$
with the servers, and generator $g_\ell$ for the given
protocol round to produce a ciphertext:
\begin{align*}
  C_i = m g_\ell^{(\sum_{j=1}^M r_{ij})}
\end{align*}
As before, $m=1$ if the sender is not the slot owner.

To prove the validity of a ciphertext element, the client 
executes the following proof of knowledge, 
where $Y$ is the slot owner's pseudonym public key,
$r_i = \sum_{j=1}^M r_{ij}$,
and $R_{ij}$ is the commitment
to the secret shared between client $i$ and server $j$:
\begin{align*}
  \textsf{PoK}\{r_{i}, y:
	((\Pi_{j=1}^M R_{ij}) = \hat{g}^{r_{i}} 
    \land 
   C_i = g_\ell^{r_{i}}) 
  \lor Y = g^y\}&
\end{align*}

\paragraph{Server Ciphertext Construction}
Server $j$'s ciphertext for the $\ell$th message exchange
round is similar to the client ciphertext,
except with negated exponents:
\begin{align*}
  S_j = g_\ell^{(- \sum_{i=1}^N r_{ij})}
\end{align*}
The server proves correctness of a ciphertext by executing
a proof of knowledge, where $r_j = \sum_{i=1}^N r_{ij}$:
\begin{align*}
  \textsf{PoK}\{r_j:
  (\Pi_{i=1}^N R_{ij}) = \hat{g}^{r_j}
  \land 
	S_j = g_\ell^{-r_j} 
  \}
\end{align*}

\paragraph{Message Reveal}
The product of the client and server ciphertexts
reveals the slot owner's plaintext message $m$: 
\begin{align*}
  m = (\Pi_{i=1}^N C_i )(\Pi_{j=1}^M S_j)
\end{align*}

\paragraph{Failed Session Set-up}

A dishonest client $i$ might try to disrupt the protocol by
publishing a corrupted commitment $R'_{ij}$ that disagrees with 
server $j$'s commitment $R_{ij}$ to the shared secret $r_{ij} = \textsf{KDF}(g^{a_i b_j})$.
If the commitments disagree, the honest server can 
prove its innocence by broadcasting the Diffie-Hellman secret
$\rho_{ij} = g^{a_i b_j}$ along with a proof that it correctly computed
the Diffie-Hellman secret using its public key $B_j$ and the client's
public key $A_i$.
\begin{align*}
  \textsf{PoK}\{b_j: \rho_{ij} = A_i^{b_j} \land B_j = g^{b_j}\}
\end{align*}
If the server is dishonest, the client can produce a similar
proof of innocence.
Any user can verify this proof, and then use $g^{a_i b_j}$
to recreate the correct commitment $R_{ij}$.
Once the verifier has the correct commitment $R_{ij}$,
the verifier can confirm either that the client
in question published an invalid commitment or that the 
server in question dishonestly accused the client.

Since the session set-up between client $i$ and server $j$ will only
fail if either $i$ or $j$ is dishonest, there is no security risk
to publishing the shared secret $g^{a_i b_j}$ after a failed set-up---%
the dishonest client (or server) could have shared 
this secret with the adversary anyway.

\paragraph{Long Messages}
The client and server ciphertext constructions 
described above allow the slot owner to transmit
a plaintext message $m$ that is at most one group 
element in length in each run of the protocol.
To encode longer plaintexts efficiently, 
participants
use a modified proof-of-knowledge construction
that proves the validity of $L$ ciphertext elements
($C_{i,1}$ through $C_{i,L}$) at once:
\begin{align*}
  \textsf{PoK}\{r_i, y: ((\Pi_{j=1}^M R_{ij} = \hat{g}^{r_i}) 
    \land (\land_{\ell=1}^L C_{i,\ell} = g_\ell^{r_i})
    ) \lor Y = g^y \}
\end{align*}
Servers can use a similarly modified proof of knowledge.
This modified knowledge proof is surprisingly compact:
the length of the proof is {\em constant} in $L$,
since the length of the proof is linear in the number of
proof variables (here, the only variables are $r_i$ and $y$).
The total length of the tuple 
$\langle \vec{C}_i, \textsf{PoK}\rangle$ 
using this proof is $L + O(1)$.

\paragraph{Lazy Proof Verification}
\label{sec:crypto-hashing-lazy}
In the basic protocol, every server verifies
the validity proof on every client ciphertext in 
every protocol round.
To avoid these expensive verification operations,
servers can use {\em lazy proof verification}:
servers check the validity of the client proofs
only if they detect, at the end of a protocol run, 
that the anonymous slot owner's message was corrupted.
For reasons discussed in 
\abbr{the extended version of this paper~\cite{corrigangibbs13proactively-tr}}{Appendix~\ref{app:lazy}},
lazy proof verification is possible only using
the pairing-based or 
hashing-generator ciphertext constructions.

\paragraph{Security Analysis}
Since the hashing-generator scheme is
the most performant variant,
we sketch an informal security proof for the hashing-generator 
proof construction in 
\abbr{the extended version of this paper~\cite{corrigangibbs13proactively-tr}}{Appendix~\ref{app:proof}}.

%% file: eval.tex
\section{Evaluation}
\label{sec:eval}

This section describes our \app prototype implementation
and summarizes the results of our evaluations.
  
\subsection{Implementation}

We implemented the \app protocol in C++ using the Qt framework
as an extension to the existing Dissent prototype~\cite{wolinsky12dissent}.
Our implementation uses OpenSSL 1.0.1 
for standard elliptic curve groups, 
Crypto++ 5.6.1 for big integer groups, 
and the Stanford Pairing-Based 
Cryptography (PBC) 0.5.12 library for
pairings~\cite{pbc}.
Unless otherwise noted, the evaluations use 1024-bit integer groups,
the 256-bit NIST P-256 elliptic curve group~\cite{nist09fips186},
and a pairing group in which $G_1$ is an elliptic curve over a 512-bit field 
(using PBC's ``Type A'' parameters)~\cite{lynn07pairing}.
We collected the macrobenchmark and end-to-end evaluation results
on the DeterLab~\cite{deterlab} testbed.

The source code for our implementation 
is available at
\url{https://github.com/DeDiS/Dissent}.

\subsection{Microbenchmarks}
\label{sec:eval-microbench}

\begin{figure}
  \centering
  \includegraphics[width=0.45\textwidth]{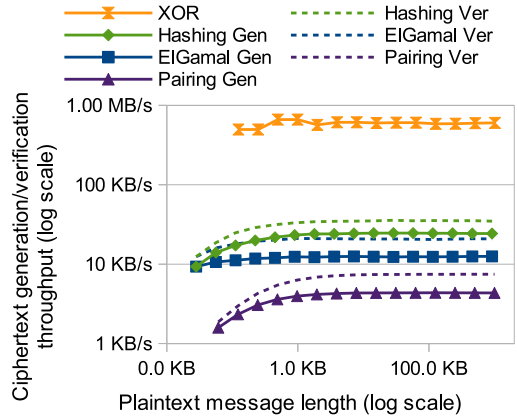}
  \caption{Ciphertext generation and 
    verification throughput for the 
    three verifiable DC-net variants and the XOR-based scheme.}
  \label{fig:eval-variant-both}
\end{figure}

To compare the pure computational costs of the different DC-net schemes,
Figure~\ref{fig:eval-variant-both} shows 
ciphertext generation and
verification throughput measured
at a variety of block sizes,
running on a workstation with a 
3.2 GHz Intel Xeon W3565 processor. 
These experiments involve no network activity, and are single-threaded,
thus they do not reflect any speedup that parallelization might offer.

The hashing-generator construction,
which is the fastest scheme tested,
encrypts 20 KB of client plaintext per second.
The slowest, paring-based construction encrypts
around 3 KB per second.  
The fastest verifiable scheme is still over an order of magnitude
slower than the traditional (unverifiable) XOR-based scheme,
which encrypts 600 KB of plaintext per second.
The hashing-generator scheme performs best because it
needs no pairing operations and
requires fewer group exponentiations than the 
ElGamal construction.

Figure~\ref{fig:eval-variant-both} shows that
ciphertext verification is slightly faster than 
ciphertext generation. 
This is because generating the ciphertext and 
zero-knowledge proof requires more
group exponentiations than proof verification does.

\begin{figure}
  \includegraphics[width=0.45\textwidth]{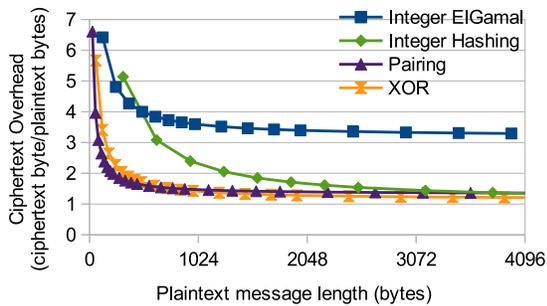}
  \caption{Ciphertext expansion factor (overhead) using the integer ElGamal-style,
    pairing-based, and hashing-generator protocol variants.}
  \label{fig:eval-overhead}
\end{figure}

The three constructions also vary
in the size of ciphertexts they generate
(Figure~\ref{fig:eval-overhead}).
While the pairing-based scheme and the 
hashing-generator schemes encrypt length $L$ plaintexts
as ciphertexts of length $L+O(1)$, the ElGamal-style scheme encrypts length $L$
plaintexts as length $3L+O(1)$ ciphertexts.
As discussed in Section~\ref{sec:crypto-elgamal}, for every plaintext message
element encrypted, ElGamal-style ciphertexts must include
an ephemeral public key and an additional 
proof-of-knowledge group element.
Since the hashing-generator scheme is the fastest
and avoids the ElGamal scheme's ciphertext expansion,
subsequent experiments use the hashing-generator scheme unless
otherwise noted.

\subsection{Accountability Cost}

\label{sec:eval-accountability}
\begin{figure*}
\includegraphics[width=\textwidth]{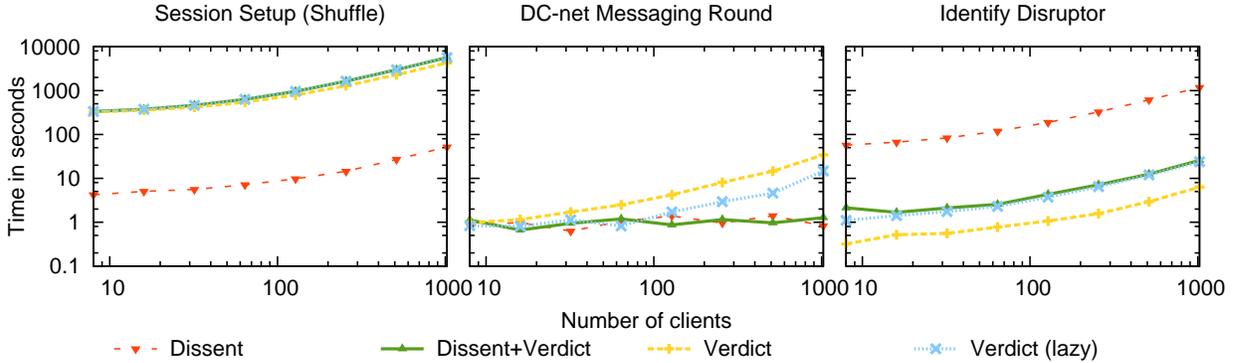}
\caption{Time required to initialize a session, perform one
  messaging round, and to identify a disruptor.}
\label{fig:macro-blame}
\end{figure*}

Figure~\ref{fig:macro-blame} presents three graphs:
(a) the time it takes to 
set up a transmission schedule via a verifiable shuffle,
prior to DC-net communication,
(b) the time required to execute a single DC-net protocol round
in each scheme, and
(c) the time required to identify a disruptor.
The graphs compare four protocol variants:
Dissent, \app, \app with lazy proof verification,
and the Dissent+\app hybrid DC-net.
We ran this experiment on DeterLab using 8 servers and 128 clients.
To scale beyond 128 clients,
we ran multiple client processes on each client machine.
Session setup time measures
the time from session start to 
just before the first DC-net messaging round.

The one-time session setup time for \app
is longer than for Dissent because the
verifiable shuffle implementation
Dissent uses is heavily optimized for shuffling DSA
signing keys. 
Shuffling \app public keys, which are drawn from
different group types, requires using 
a less-optimized version of the verifiable shuffle.
We do not believe this cost is fundamental to the \app approach,
and in any case these setup costs are typically amortized
over many DC-net rounds.

The Dissent+\app hybrid DC-net is just as fast as Dissent
in the normal case,
since Dissent and the hybrid DC-net run {\em exactly the same code}
if there is no active disruptor in the group.
Network latency comprises the majority of the time for
a messaging round when using the Dissent and the hybrid 
Dissent+\app DC-nets---%
messaging rounds take between 0.6 and 1.4 seconds to complete
in network sizes of 8 to 1,024 clients.
In contrast, \app becomes computationally
limited at 64 clients, taking approximately 2.5 seconds per round.
\App (lazy) improves upon this by becoming computationally
limited at 256 clients, requiring approximately 3 seconds per messaging
round.

\App incurs the lowest accountability (blame) cost
of the four schemes.
\App's verifiable DC-net checks the validity of 
each client ciphertext before processing it further,
so the time-to-blame in \app is equal to the cost 
of verifying the validity proofs on $N$ client ciphertexts.
``\App (lazy)'' uses the lazy proof verification
technique described in Section~\ref{sec:crypto-hashing}---%
servers verify the client proofs of correctness 
only if they detect a disruption.
Lazy proof verification delays the verification operation
to the end of a messaging phase, so the time-to-blame
is slightly higher than in pure \app.

Dissent, which has the highest time-to-blame,
has an accountability process that requires 
the anonymous client whose message was corrupted to
submit an ``accusation'' message to 
a lengthy verifiable shuffle protocol,
in which all members participate.
This verifiable shuffle is the reason that Dissent
takes the longest to identify a disruptor.
The hybrid Dissent+\app DC-net (Section~\ref{sec:proto-hybrid}) 
avoids Dissent's extra verifiable shuffle by 
falling back instead to a verifiable DC-net to resolve disruptions.

As Figure~\ref{fig:macro-blame} shows, 
the messaging round time in the hybrid Dissent+\app DC-net 
is as fast as in Dissent, but the hybrid scheme
reduces Dissent's time to detect misbehavior by roughly two
orders of magnitude.

\subsection{Anonymous Microblogging}
\label{sec:eval:twitter}

\App's ability to tolerate many dishonest nodes
makes it potentially attractive for anonymous microblogging
in groups of hundreds of nodes.
In Twitter, messages have a maximum length of 140 bytes,
which means that a single tweet can fit into a 
few 256-bit elliptic curve group elements.
Twitter users can also tolerate messaging 
latency of tens of seconds or even a few minutes,
which would be unacceptable for interactive web browsing.

This experiment evaluates the suitability of \app 
for {\em small-scale} anonymous microblogging applications,
giving users anonymity among hundreds of nodes,
e.g., for students microblogging on a university campus.
To test \app in this scenario,
we recorded 5,000 Twitter users' activity for one-hour
and then took subsets of this trace: 
the smallest subset contained only the Tweets of the 40 most 
active users, and the largest subset contained the Tweets
of the 1,032 most active users.
We replayed each of these traces through Dissent and through
\app, using each of the three ciphertext constructions.

\begin{figure}
  \includegraphics[width=0.45\textwidth]{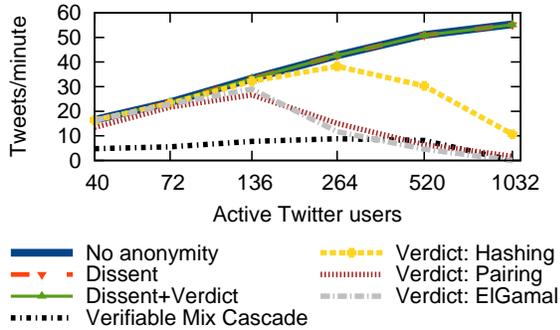}
  \caption{Rate at which various anonymity schemes process tweets,
	for varying numbers of active users.}
  \label{fig:twitter-bw}
\end{figure}

We ran our experiment on DeterLab~\cite{deterlab},
on a test topology consisting of
eight servers connected to a 100 Mbps LAN with 10 ms of
server-to-server latency, and with each set of clients 
connecting to their upstream server over 
a {\em shared} 100 Mbps link with 50 ms of latency.
Scarcity of testbed resources limited the number of 
available delay links, but our experiment attempts to
approximate a wide-area deployment model in which clients
are geographically dispersed and bandwidth-limited.

Figure~\ref{fig:twitter-bw} shows the 
Tweet-rate latency induced by the different
anonymity systems relative to the baseline (no anonymity)
as the number of active users (and hence, the anonymity set size)
in the trace increases.
Both Dissent and the Dissent+\app hybrid systems can
keep pace with the baseline in a 1,000-node network---%
the largest network size feasible on our testbed.
The pure \app variants could not keep pace with the baseline
in a 1,000-node network, while hashing-generator 
variant of \app runs almost as quickly as the 
baseline in an anonymity set size of 264.
These results suggest that \app might realistically 
support proactively
accountable anonymity for microblogging groups of 
up to hundreds of nodes.

Figure~\ref{fig:twitter-bw} also compares
\app to a mix-net cascade (a set of mix servers) in which 
each mix server uses a Neff proof-of-knowledge~\cite{neff01verifiable} to demonstrate
that it has performed the mixing operation properly.
Like \app, this sort of mix cascade forms
a traffic-analysis-resistant anonymity system,
so it might be used as an alternative to \app for anonymous messaging.
Our evaluation results demonstrate that the hashing-generator
variant of \app outperforms the mix cascade at all network sizes
and that the Tweet throughput of the Dissent+\app hybrid is more than
$6\times$ greater than the throughput of the mix cascade at a network
size of 564 participants.

\subsection{Anonymous Web Browsing}

Dissent demonstrated that accountable DC-nets are fast enough
to support anonymous interactive Web browsing
in local-area network deployments~\cite{wolinsky12dissent}.
We now evaluate whether \app is similarly usable in a web browsing scenario.
Our experiment simulates a group of
nodes connected to a single WLAN network.
This configuration emulates, for example, a
group of users in an Internet caf\'e 
browsing the Internet anonymously.

In our simulation on DeterLab~\cite{deterlab}, 
8 servers and 24 clients communicate over a
network of 24 Mbps links with 20 ms node-to-node latency.
To simulate a Web browsing session,
we recorded the sequence of requests and responses that
a browser makes to download home page content 
(HTML, CSS files, images, etc.)
from the Alexa ``Top 100'' Web pages~\cite{alexa}.
We then replayed this trace with the client using
one of four anonymity overlays: no anonymity, the Dissent DC-net,
the \app-only DC-net, and the Dissent+\app hybrid DC-net.
The simulated client sends the upstream
(request) traffic through the anonymity network and
servers broadcast the downstream (response) traffic 
to all nodes.

\begin{figure}[t]
  \includegraphics[width=0.45\textwidth]{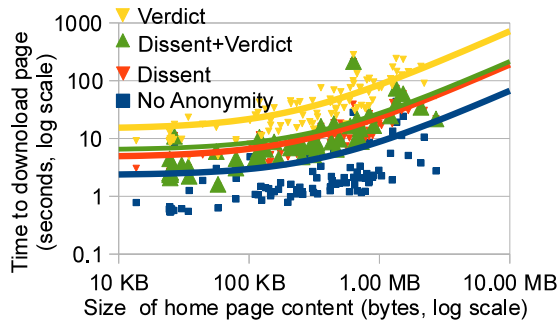}
  \caption{Time required to download home page context for
    Alexa ``Top 100'' Web sites (with linear trend lines).}
  \label{fig:macro-top100}
\end{figure}

\begin{figure}[t]
  \includegraphics[width=0.45\textwidth]{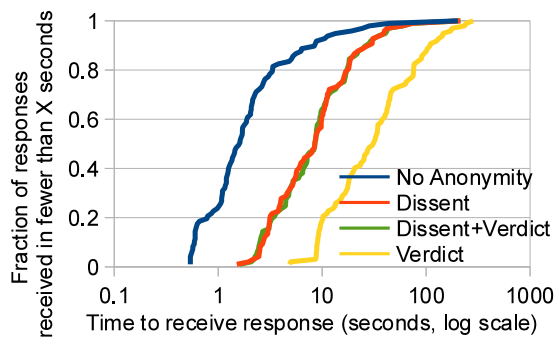}
  \caption{CDF of time required to download home page context
  for Alexa ``Top 100'' Web sites.}
  \label{fig:macro-top100-cdf}
\end{figure}

Figure~\ref{fig:macro-top100} charts the time required
to download all home page content using the four different
network configurations.
The median Web page took one
second to load with no anonymity, fewer than
10 seconds over Dissent, and around 30 seconds using
\app only (Figure~\ref{fig:macro-top100-cdf}).
Notably, the hybrid Dissent+\app scheme exhibits performance
nearly identical to that of Dissent alone, since it 
it falls back to the slower verifiable \app DC-net
only when there is active disruption.
The \app-only DC-net is much slower than Dissent
because every node must generate a computationally expensive
zero-knowledge proof in every messaging round.

These experiments show that \app adds
no overhead to Dissent's XOR-based DC-net
in the absence of disruption.
In addition, these experiments illustrate the flexibility
of verifiable DC-nets,
which can be used either as a ``workhorse'' for anonymous communication
or more selectively in combination with traditional XOR-based DC-nets;
we suspect that other interesting applications 
will be discovered in the future.

%% file: rel.tex
\section{Related Work}
\label{sec:rel}

Chaum recognized the risk of anonymous disruption attacks
in his original formulation of DC-nets~\cite{chaum88dining},
and proposed a probabilistic tracing approach based on {\em traps},
upon which Waidner and Pfitzmann expanded~\cite{waidner89dining}.

Herbivore~\cite{sirer04eluding,goel03herbivore}
sidestepped the disruption issue by forming groups dynamically,
enabling nodes to leave disrupted groups and form new groups
until they find a disruption-free group.
Unfortunately, the likelihood that a group contains some malicious node
likely increases rapidly with group size, and hence anonymity set,
limiting this and related partitioning approaches~\cite{ahn03kanonymous}
to systems supporting small anonymity sets.
Further, in an analog to a known attack against Tor~\cite{borisov07denial},
an adversary might selectively disrupt only groups he has
only {\em partially} but not {\em completely} compromised.
With a powerful adversary controlling many nodes,
after some threshold a victim becomes {\em more} likely to ``settle into''
a group that works precisely because it is {\em completely} compromised,
than to find a working uncompromised group.

$k$-anonymous message transmission~\cite{ahn03kanonymous} also achieves 
disruption resistance by partitioning participants into small disruption-free groups. 
A crucial limitation of the $k$-anonymity system is that an honest
client is anonymous {\em only} among a small constant ($k$)
number of nodes.
In contrast, \app clients in principle obtain anonymity among the set of
{\em all} honest clients using the system.

Dissent~\cite{corrigangibbs10dissent,wolinsky12dissent}
uses verifiable
shuffles~\cite{brickell06efficient,neff01verifiable}
to establish a {\em transmission schedule} for DC-nets,
enabling groups to guarantee a one-to-one correspondence of
group members to anonymous transmission slots.
The original Dissent protocol~\cite{corrigangibbs10dissent}
offered accountability but limited performance.
A more recent version~\cite{wolinsky12dissent}
improves performance and scalability,
but uses a retrospective ``blame'' protocol which
requires an expensive shuffle when disruption is detected.

Golle and Juels~\cite{golle04dining}
introduced the verifiable DC-net concept
and formally developed a scheme based on bilinear maps,
forming \app's starting point.
To our knowledge this scheme was never implemented
in a working anonymous communication system, however,
and we find that its expensive pairing operations
limit its practical performance.

Crowds~\cite{reiter99crowds},
LAP~\cite{hsiao12lap},
Mixminion~\cite{danezis03mixminion},
Tarzan~\cite{freedman02tarzan}, and
Tor~\cite{dingledine04tor}, 
provide anonymity in large networks, 
but these systems cannot protect
against adversaries that observe traffic~\cite{bauer07low,murdoch07sampled} 
or perform active attacks~\cite{borisov07denial} on
a large fraction of network links.
\App maintains its security properties in the
presence of this type of strong adversary.
A cascade of cryptographically verifiable 
shuffles~\cite{furukawa01efficient,neff01verifiable}
can offer the same security guarantees that \app does, 
but these shuffles generally require more expensive proofs-of-knowledge.

%% file: concl.tex
\section{Conclusion}
\label{sec:concl}

\App is a new anonymous group messaging system
that combines the traffic analysis resistance of DC-nets
with disruption resistance based on public-key cryptography
and knowledge proofs.
Our experiments show that \app may be suitable for messaging
in groups of hundreds to thousands of users,
and can be combined with traditional XOR-based DC-nets
to offer good normal-case performance
while reducing the system's vulnerability to disruption events
by two orders of magnitude.

\subsection*{Acknowledgments}

We wish to thank 
Aaron Johnson,
Ewa Syta, 
Michael J. Fischer,
Michael Z. Lee,
Michael ``Fitz'' Nowlan, and
Ramki Gummadi for their helpful comments.
We also thank our shepherd, Micah Sherr,
and the anonymous USENIX reviewers,
for their valuable feedback.
Finally, we thank the DeterLab staff for 
their flexibility, patience, and support during 
the evaluation process.
%
%
%
This material is based upon work supported by the Defense Advanced Research
Agency (DARPA) and SPAWAR Systems Center Pacific, Contract No. N66001-
11-C-4018.
%
%

%% file: malkeys.tex
\section{Maliciously Crafted Public Keys}
\label{app:malkeys}

The protocol construction does {\em not} assume 
the existence of a trusted third party who generates
participants public/private keypairs.
Instead, every client must verify the validity
of every server's public key $g^a$ at
the start of a protocol run, and every 
server must verify the validity of every client's
public key as well.

Participants can verify that a public key $g^a$ is
valid by confirming that $g^a \in G$.
However, this check is not enough to prevent malicious
participants from disrupting a protocol run.
For example, if an honest server published a public
key $A = g^a$, a dishonest server could publish a
public key $B = (g^b)(g^a)^{-1} = g^{b-a}$ whose
corresponding secret key is a function
of $A$'s secret key. 
If a client then uses the ElGamal-style 
ciphertext construction to 
create a ciphertext using ephemeral private key 
$r$ encrypted for both servers,
the product of the two servers' public keys
will result in an encryption that server $B$ can
decrypt unilaterally:
\begin{align*}
  C = m(AB)^{r} = m((g^a)(g^b)(g^a)^{-1})^r = mg^{rb}
\end{align*}
By creating a public key in this way, server $B$
can ``cancel out'' the effect of $A$'s public key
even without $A$'s cooperation.

To prevent this sort of attack, we require participants
to prove knowledge of the discrete logarithm of their 
public key.
In the notation of Camenisch and Stadler, participant
must prove:
\begin{align*}
  \textsf{PoK}\{a: A = g^a\}
\end{align*}
Requiring this proof defeats the attack outline above,
because without knowledge of $A$'s secret key $a$
server $B$ cannot prove knowledge of the secret exponent
$b-a$.
Since we use public keys of the form $g^a$ in each of
the three ciphertext constructions,
it is possible to execute this proof of knowledge
regardless of which scheme is in use.

%% file: protosec.tex
\section{Security of Messaging Protocol}
\label{app:protosec}

This section sketches a security argument 
that the messaging protocol described
in Section~\ref{sec:proto} 
satisfies the security properties
of integrity, anonymity, and accountability,
provided that the underlying cryptographic primitives 
(described in Section~\ref{sec:crypto}) are correct.

\subsection{Integrity}

Since every protocol message is signed with the
sender's long-term signing key, and since each message
includes a round nonce unique to this particular run of the
protocol, replay and impersonation attacks are infeasible.

Assume, by way of contradiction,
that some honest client $h$ concludes that the protocol
run has terminated in success {\em and} that $h$ holds a final
output message $m'$ that is unequal to the slot owner's
plaintext message $m$.

For $h$ to conclude that the protocol run succeeded, $h$ would
have had to receive a valid signature on the invalid ciphertext
$m'$ from each of the $M$ servers in Phase~\ref{proto:client-ver}

Since at least one of the $M$ servers is honest, at least one
of the servers (call this server $s$), verified each of 
the proofs of correctness on each of the server ciphertexts
in Phase~\ref{proto:server-plaintext-reveal}.
Since (by assumption) the ciphertext construction is sound,
each of the server ciphertexts must therefore be a
correctly formed server ciphertext (except with negligible probability) 
corresponding to the client ciphertext set that $s$ received
in Phase~\ref{proto:server-plaintext-share}.

In addition, honest server $s$ has verified the proof of correctness on
each of the client ciphertexts (Phase~\ref{proto:server-ciphertext-gen}), 
and the ciphertext soundness property (Section~\ref{sec:crypto-prim})
means that each of these ciphertexts is also validly constructed.

Since the plaintext element $m'$ that $h$ receives 
is the product of $N$ valid client ciphertexts,
and of $M$ valid server ciphertexts---one from each server and constructed
in response to the client ciphertext set---the product of these ciphertexts
will be the slot owner's original plaintext $m$.
Since we have $m = m'$, but $m \neq m'$ by assumption, this is
a contradiction.

\subsection{Anonymity}

The anonymity of the overall protocol derives directly from the anonymity
of the underlying cryptographic construction (Section~\ref{sec:crypto}).
To break the anonymity of the system, an adversary must gain some 
advantage in distinguishing the slot owner from other
clients participating in the protocol.
Since the only difference in behavior between the slot owner
and the remaining clients comes in Phase~\ref{proto:client-ciphertext-gen}
of the protocol (in which participants generate their ciphertext
messages), the adversary must be able to use the ciphertext messages
alone to distinguish the slot owner from others.
However, since we assume that the underlying cryptographic primitive
maintains slot owner anonymity, the attacker has no feasible way to
distinguish the slot owner's ciphertext from the remaining
ciphertexts.
We discuss the specific indistinguishability assumptions used
in each cryptographic construction in Section~\ref{sec:crypto}.

\subsection{Accountability}
\label{app:protosec-accountability}

The accountability property requires that if an honest
node concludes that a protocol run has terminated in failure,
it holds a third-party verifiable proof of (at least) one
dishonest node's misbehavior.

We enumerate the ways in which a node can misbehave and demonstrate
how an honest node can detect each type of misbehavior:
\begin{itemize}

  \item \textbf{Invalid public key (session set-up)}
        A client or server who submits an invalid public key
        or an invalid proof of knowledge during the session
        set-up process will be immediately exposed by the recipient, 
        and the message containing the invalid key becomes proof of
        the key generator's dishonesty.

        The proof of knowledge ensures that a dishonest node
        will be unable to pass off an honest node's public key
        as his own (since the dishonest node will be unable to produce
        a signature of his own long-term public key with the
        honest node's public key).

  \item \textbf{Invalid client ciphertext (Phase~\ref{proto:client-ciphertext-gen})}
        A client who submits an invalid ciphertext in 
        Phase~\ref{proto:client-ciphertext-gen} will be exposed
        by the receiving server in Phase~\ref{proto:server-ciphertext-share}
        (if the receiving server is honest) when the server
        checks the validity of the client's ciphertext.
        The signed, invalid ciphertext submitted by the client
        is the proof of the client's dishonesty.
        If the {\em server} is dishonest, the server
        will be exposed later in the protocol.

  \item \textbf{Client equivocation (Phase~\ref{proto:client-ciphertext-gen})}
        A client who submits two different ciphertexts to two
        different servers will be exposed by an honest server
        in Phase~\ref{proto:server-ciphertext-gen} when the
        server observes the two different valid signed client ciphertexts
        in two servers' client ciphertext sets.
        These two ciphertexts will become the proof of the client's
        dishonesty.

  \item \textbf{Server accepts invalid ciphertext 
        (Phase~\ref{proto:server-ciphertext-share})}
        If a server accepts an 
        {\em invalid} ciphertext from a client, an honest server
        will detect this dishonesty in Phase~\ref{proto:server-ciphertext-gen},
        since the dishonest server's ciphertext set will not match the
        honest server's ciphertext set (since honest servers will only 
        transmit a ciphertext set if all client ciphertexts are valid).
        The honest server will use the dishonest server's ciphertext
        set as its proof of the server's dishonesty.

  \item \textbf{Invalid server ciphertext (Phase~\ref{proto:server-ciphertext-gen})}
        An honest server in Phase~\ref{proto:server-plaintext-reveal} will expose
        a dishonest server that has broadcasted an invalid server ciphertext 
        in Phase~\ref{proto:server-ciphertext-gen}.
        The invalid server ciphertext serves as the proof of the server's dishonesty.

  \item \textbf{Invalid server signature (Phase~\ref{proto:server-plaintext-reveal})}
        Since, in Phase~\ref{proto:server-plaintext-reveal}, all servers hold 
        the same set of valid client ciphertexts, and a set of $M$ valid server
        ciphertexts, each honest server will be able to recover the same
        plaintext message.
        An honest server in Phase~\ref{proto:server-plaintext-share} will
        expose a dishonest server that has broadcasted an invalid signature
        in Phase~\ref{proto:server-plaintext-reveal} when the honest server
        checks the dishonest server's signature against the revealed plaintext.
        The invalid signature serves as proof of the server's dishonesty.

  \item \textbf{Corrupted signature from server (Phase~\ref{proto:server-plaintext-share})}
        A server that transmits invalid signatures to its downstream clients
        in Phase~\ref{proto:server-plaintext-share} will be revealed by honest
        clients in Phase~\ref{proto:client-ver}.
        Since an honest server will verify each signature before sending them
        to its downstream clients, an honest client can conclude that if any
        of the signatures on the plaintext is invalid, its upstream server
        is dishonest.
        
  \item \textbf{Un-parseable message (all phases)} 
        If an honest participant ever receives an un-parseable 
        message from another participant, that carries a valid signature with
        the sender's long-term signing key, the signed message becomes
        proof of the dishonest node's misbehavior.

\end{itemize}

%% file: proof.tex
\section{Security Arguments}
\label{app:proof}

This section provides a game-based definition 
of the anonymity property (introduced in Section~\ref{sec:crypto-prim})
and then argues for the security of the hashing-generator
ciphertext construction (introduced in Section~\ref{sec:crypto-hashing}).

\subsection{Anonymity Game}
We say that a protocol maintains {\em slot owner anonymity}
if the advantage of any polynomial-time adversary in the
following anonymity game is negligible (in the implicit security parameter).
The game, which takes place between an adversary and a
challenger, proceeds as follows:
\begin{itemize}
  \item The challenger picks per-session keypairs for the slot owner,
        for each of the $M$ servers, 
        and for each of the $N$ clients.
  \item The challenger sends to the adversary:
        \begin{itemize}
          \item all of the public keys, 
          \item the private keys for the $M-1$ dishonest servers, and
          \item the private keys for the $N-2$ dishonest clients.
        \end{itemize}
        The challenger holds the private key for the one honest server
        and the two honest clients.

  \item The adversary picks a plaintext message $m$ and sends it to
        the challenger.
  \item The challenger picks a value $\beta \in \{1, 2\}$ at random.
        The challenger sets the slot owner to be honest client $\beta$.
  \item The challenger and adversary run the anonymous communication 
        protocol, with the challenger playing the role of the honest
        participants and the adversary playing the role of the dishonest
        participants. 
  \item At the conclusion of the protocol, the adversary makes a guess
        $\beta'$ of the value $\beta$.
\end{itemize}
The adversary's advantage $\epsilon$ in the game 
is $|\textrm{Pr}[\beta = \beta'] - \frac{1}{2}|$.
If $\epsilon$ is negligible, we say that the 
protocol maintains {\em slot owner anonymity}.

Note that this formulation of the anonymity game 
assumes that all participants' keypairs are generated
honestly (i.e., that a dishonest node's public key
does not depend on an honest node's public key in 
a way that could harm the security of the protocol).
In practice, we assure that this is true using a
proof of knowledge, described in Appendix~\ref{app:malkeys}.

\subsection{Hashing-Generator Scheme}

In the following section, we demonstrate that
in the random oracle model~\cite{bellare93random}
and under the decision Diffie-Hellman
assumption~\cite{boneh98decision},
the hashing-generator ciphertext construction
satisfies the security properties of Section~\ref{sec:crypto-prim}.
The security arguments for the ElGamal-style scheme and the pairing-based
schemes proceed in a similar fashion, but we focus on the 
hashing-generator construction because it is the most performant
of the three variants.

\paragraph{Proof Sketch (Completeness, Soundness, Zero-Knowledge)}
The client and server ciphertexts are 
non-interactive zero-knowledge proofs of knowledge,
adopted directly from prior 
work~\cite{camenisch97proof,cramer94proofs,fiat87prove,groth04honest}.
Our completeness, soundness, and zero-knowledge properties
of Section~\ref{sec:crypto-prim} follow immediately from the 
completeness, special soundness, and special honest-verifier 
zero knowledge properties of the underlying proof system.

\paragraph{Proof Sketch (Integrity)}
Since the ciphertext construction maintains soundness
(as argued above), any client or server ciphertext
that an honest node finds to be valid 
will have the correct form.
Taking the product of $N$ correctly 
constructed client ciphertexts and $M$ 
correctly constructed server ciphertexts will
result in the cancellation of each pair
of client/server shared secrets $g_\ell^{r_{ij}}$
in the product.
(This is because both $g_\ell^{r_{ij}}$ and its
inverse are included in the product.)
The resulting product, then, will be
the slot owner's plaintext message $m$.

\paragraph{Proof Sketch (Anonymity)}
An adversary who wants to break the protocol's anonymity
has two choices: (1) the adversary can deviate from the protocol
specification in some malicious way, or (2) the adversary 
can follow the protocol specification exactly and try to 
break the anonymity by passively observing messages from 
honest nodes.

Consider the adversary's first option (to violate the protocol in
a malicious way).
The only opportunities that an adversarial client has to deviate
from the protocol are to 
(a) publish an incorrect commitment to its shared secret or
(b) submit an invalid client ciphertext.
Option (a) will not help a malicious client, since the 
invalid commitment will be detected by honest nodes and will halt
the protocol.
Option (b) will not help a malicious client either, because
the soundness property of the zero-knowledge proof system means
that an invalid ciphertext will be rejected by the honest
server (thereby halting the protocol run).

An adversarial server can also violate the protocol,
but these violations will not help it win the anonymity game either.
The server can publish an invalid commitment in the setup phase, but 
honest nodes will detect this.
The server can try to manipulate the set of client ciphertexts, but
all honest nodes will detect this as well.
The server can broadcast an invalid server ciphertext, but honest
nodes will detect this also.

The accountability property of the messaging 
protocol, in combination with the soundness property of the
zero-knowledge proof construction, makes it infeasible for
the adversary to gain any information by violating the
protocol.
(Of course, the adversarial nodes can try to collude
to disrupt the protocol as well, but collusion will not
enable the adversary to learn anything that it does not
already know.)

The adversary's other possible strategy 
is to follow the protocol correctly and to try to 
guess who the slot owner is based on a successful run of
the protocol.
We demonstrate that this strategy is ineffective as well:
if there exists an efficient algorithm
$A$ that has an advantage in the anonymity game, there is
another efficient algorithm $B$ that has an advantage in
the decision Diffie-Hellman (DDH) game, contradicts
the DDH assumption.

The input to the algorithm $B$ is a DDH challenge tuple
$\langle \hat{g}, \hat{g}^x, \hat{g}^y, \hat{g}^z \rangle$ and $B$ must output
``yes'' if $z = xy$ and ``no'' otherwise.
(This generator $\hat{g}$ is the same generator
 $\hat{g}$ used in Section~\ref{sec:crypto-hashing}.)

To use algorithm $A$ as a subroutine, $B$ 
must simulate $A$'s view of a run of the protocol.
The simulation proceeds as follows, with $B$ simulating 
the role of the challenger
and with $A$ playing the role of the adversary:
\begin{itemize}

  \item The simulator decides which honest client
        $\beta \in \{1, 2\}$ will serve as the anonymous slot owner.
        (We describe the case in which $\beta = 1$, but if
        $\beta = 2$, the roles of the two honest clients are simply swapped.)

  \item The simulator selects public keys for all of the
        clients and servers. 
        In the setup phase, the simulator programs the 
        key-derivation function \textsf{KDF}, modeled as a random oracle,
        to (virtually) assign the value $y$ to the
        the secret shared $r_{1,1}$ between honest client
        $h_1$ and honest server $s_1$. 
        (We say that $r_{1,1}$ is ``virtually'' set to $y$ because
        the simulator does not actually know the value $y$.
        The simulator only needs to reveal $\hat{g}^y$ to $A$,
        so the simulator never needs to know $y$ itself.)
        The simulator assigns random values to all 
        secrets $r_{ij}$ shared between every other
        client/server pair.

  \item The simulator must publish commitments $R_{ij} = \hat{g}^{r_{ij}}$ 
        to each of the honest participants' shared secrets.
        The simulator knows all of the values $r_{ij}$ (except $r_{1,1}$),
        so it can compute almost all of these commitments directly.
        The simulator uses $\hat{g}^y$ from the DDH challenge as the
        commitment to $r_{1,1}$.

  \item The simulator receives a message $m$ from $A$.

  \item Recall that in the hashing-generator scheme, all participants
        use a public hash function $\textsf{H}$ (again modeled as a
        random oracle) to select the group generator used in each 
        message transmission round.
        
        The simulator programs $\textsf{H}$ to output $\hat{g}^x$ 
        as the generator in the first transmission round.
        The simulator uses $\hat{g}^z$ from the DDH challenge
        tuple to simulate the honest clients' ciphertexts:
        \begin{align*}
          C_1 &= m (\hat{g}^z) (\hat{g}^x)^{\Sigma_{j=2}^M r_{1,j}} \\
          C_2 &= (\hat{g}^x)^{\Sigma_{j=1}^M r_{2,j}}
        \end{align*}
        Similarly, the simulated server ciphertext is:
        \begin{align*}
          S_1 &= (\hat{g}^z)^{-1}(\hat{g}^x)^{-\Sigma_{i=2}^N r_{i 1}} 
        \end{align*}

        Recall that both the client and server ciphertexts must
        carry non-interactive zero-knowledge proofs of correctness.
        In the random-oracle model, the simulator can 
        efficiently simulate these proofs using standard techniques~\cite{bellare93random}
        (i.e., by picking the ``challenge'' value used in the
        proof before generating the proof's commitments).

\end{itemize}

At the conclusion of the simulation, algorithm $A$ will output
a guess $\beta' \in \{1, 2\}$ that honest client $h_{\beta'}$ was the slot owner
during the protocol run. 
Algorithm $B$ outputs ``yes'' (the challenge tuple is a Diffie-Hellman tuple)
if $\beta' = 1$ and ``no'' otherwise.

With probability $1/2$, (when $z = xy$), $B$ will correctly 
simulate the view of the challenger and $B$ will win the
DDH game with probability $\epsilon$.
With probability $1/2$, (when $z \neq xy$), $B$ will produce
ciphertexts $C_1$ and $C_2$ that are group elements selected
at random by the DDH challenger.
In this latter case, $B$ will have no advantage in the DDH game
(since $A$ cannot possibly have an advantage in the anonymity game).

Thus, if $A$'s advantage in winning the anonymity game is some
non-negligible value $\epsilon$, then
$B$'s advantage in winning the DDH game is $\epsilon/2$ (which is non-negligible).


%% file: zkp.tex
\section{Zero-Knowledge Proof Instantiation}
\label{app:zkp}

The cryptographic constructions presented in 
Section~\ref{sec:crypto} make extensive use of
non-interactive zero-knowledge proofs of knowledge.
This section presents an example instantiation of one
such proof of knowledge to make the
technique concrete.
The proof-of-knowledge techniques used in \app follow
primarily from the work of Camenisch and Stadler~\cite{camenisch97proof}
which follows in turn from earlier work on proofs
of knowledge~\cite{cramer94proofs,fiat87prove,schnorr91efficient}.

Proof-of-knowledge protocols are interactive by nature,
but they can be made non-interactive by replacing the
role of an interacting verifier with a hash function.
This technique, developed by Fiat and Shamir~\cite{fiat87prove},
allows for security proofs in the ``random-oracle model''~\cite{bellare93random}.
To avoid the unwieldy phrase ``non-interactive honest-verifier 
computationally zero-knowledge proof of knowledge,'' we write ``proof.''

Our construction uses non-interactive proofs based on Schnorr's
proof of knowledge of discrete logarithms~\cite{schnorr91efficient},
proof of equality of discrete logarithms~\cite{chaum92wallet},
and witness-hiding proofs~\cite{cramer94proofs} 
(which demonstrate that the prover 
knows at least one out of $n$ secrets).

We will demonstrate an instantiation of the client
ciphertext proof of correctness used in the ElGamal-style
construction, introduced in Section~\ref{sec:crypto-elgamal}.
Proofs for the other ciphertext constructions have almost
identical form, so we omit their description herein.

This description assumes that all group elements
are elements of a group $G$ of order $q$ such that the
decision Diffie-Hellman problem~\cite{boneh98decision} is hard in $G$.
Given an ephemeral public key $R=g^r$, a ciphertext element $C$,
server public keys $\langle B_1, \dots B_M \rangle$,
and the slot owner's pseudonym public key $Y$, the client
executes a proof of knowledge over the secrets $r$ (the
ephemeral secret key) and $y$ (the slot owner's pseudonym secret key).
If the client is the slot owner, the client will know the pseudonym
secret key $y$.
If the client is {\em not} the slot owner, the client will know
the ephemeral secret key $r$.
The proof has the form:
\begin{align*}
  \textsf{PoK}\{r, y:
    (R = g^r \land C = (\Pi_{j=1}^M B_j)^r )
    \lor Y = g^y \}
\end{align*}

To simplify the notation, we relabel the variables such 
that each discrete logarithm relationship has the 
form $y_i = g_i^{x_i}$.
Note that the $g$ and $y$ values are public, while
the $x$ values are known only to the prover (in this case, 
prover is the client generating the ciphertext).
\begin{align*}
  g_1 &= g & x_1 &= r & y_1 &= R\\ 
  g_2 &= \Pi_{j_1}^M B_j & x_2 &= r & y_2 &= C\\
  g_3 &= g & x_3 &= y & y_3 &= Y\\
\end{align*}

To rewrite the proof statement with the new variable names:
\begin{align*}
  \textsf{PoK}\{x_1, x_2, x_3:
    (y_1 = g_1^{x_1}
     \land
     y_2 = g_2^{x_2}
     \land x_1 = x_2) 
  \lor
    y_3 = g_3^{x_3}
  \}
\end{align*}

The proof has three phases: 
commitment, challenge, and response.
Application of the Fiat-Shamir heuristic~\cite{fiat87prove}
means that interaction with the verifier in the
``challenge'' phase is replaced by a call to a
hash function $\mathcal{H}$ (modeled as a random oracle).

There are two versions of this proof:
one in which the prover knows $x_1$ 
and $x_2$ (recall that $x_1 = x_2$),
and another in which the prover knows $x_3$.
These proofs have similar structure, so we present
only the former variant.

\paragraph{Commitment}
The commitment values $(t_1, t_2, t_3)$ are:
\begin{align*}
  \begin{aligned}
    v_1, v_2, w &\in_R \mathbb{Z}_q&
  \end{aligned}\\
  \begin{aligned}
    t_1 &= g_1^{v_1} &
    t_2 &= g_2^{v_1} &
    t_3 &= y_3^{w} g_3^{v_2}
  \end{aligned}
\end{align*}
\paragraph{Challenges}
The challenge values $(c_1, c_2)$ are:
\begin{align*}
  h &= \mathcal{H}(g_1, g_2, g_3, y_1, y_2, y_3, t_1, t_2, t_3)\\ 
  c_1 &= h - w \mod q\\
  c_2 &= w
\end{align*}
\paragraph{Response}
The response values $(r_1, r_2)$ are:
\begin{align*}
  r_1 &= v_1 - c_1 x_1 \mod q\\
  r_2 &= v_2
\end{align*}
The final proof is $(c_1, c_2, r_1, r_2)$.

\paragraph{Verification}

To verify the proof, the verifier 
first recreates the commitments:
\begin{align*}
  t'_1 &= y_1^{c_1} g_1^{r_1}&
  t'_2 &= y_2^{c_1} g_2^{r_1}&
  t'_3 &= y_3^{c_2} g_3^{r_2}
\end{align*}

If the proof is valid, then the $t'$s values should be
equal to the original $t$ values
(recall that $x_1 = x_2$):
\begin{align*}
  \begin{aligned}
  t'_1 &= y_1^{c_1} g_1^{r_1}\\
       &= y_1^{h - w} g_1^{v_1 - c_1 x_1}\\
       &= (g_1^{x_1})^{h - w} g_1^{v_1 - c_1 x_1}\\
       &= g_1^{h x_1 - w x_1 + v_1 - (h - w) x_1} \\
       &= g_1^{h x_1 - w x_1 + v_1 - h x_1 + w x_1}\\
       &= g_1^{v_1}\\
       &= t_1\\
  \end{aligned}\qquad
  \begin{aligned}
  t'_2 &= y_2^{c_1} g_2^{r_1}\\
       &= y_2^{h - w} g_2^{v_1 - c_1 x_1}\\
       &= (g_2^{x_2})^{h-w} g_2^{v_1 - c_1 x_2}\\
       &= g_2^{h x_2 - w x_2 + v_1 - (h - w) x_2}\\
       &= g_2^{h x_2 - w x_2 + v_1 - h x_2 + w x_2}\\
       &= g_2^{v_1}\\
       &= t_2\\
  \end{aligned}
\end{align*}
\begin{align*}
  t'_3 &= y_3^{c_2} g_3^{r_2}\\
       &= y_3^{w} g_3^{v_2}\\
       &= t_3
\end{align*}

Finally, the verifier confirms
that $c_1 + c_2 \stackrel{?}{=} \mathcal{H}(g_1, g_2, g_3, y_1, y_2, y_3, t_1, t_2, t_3)$.

\paragraph{Security}

The protocol described above satisfies 
standard security properties for proof-of-knowledge
protocols~\cite{cramer94proofs}.
We briefly summarize these properties and sketch a proof
that they hold for this proof-of-knowledge.

\begin{itemize}
  \item \textbf{Completeness}: an honest verifier always accepts
        a proof generated by an honest prover.
        This is verified by simply observing (by algebraic manipulation) 
        that any valid proof generated by the prover will satisfy the
        verification equations.

  \item \textbf{Special soundness}:
        given any valid pair of commit-challenge-response transcripts
        $(t_1, t_2, t_3, c_1, c_2, r_1, r_2)$ and
        $(t_1, t_2, t_3, c'_1, c'_2, r'_1, r'_2)$ such that
        $c_1 \neq c'_1$ and $c_2 \neq c'_2$,
        it is possible to extract the prover's secret value used
        to generate the proof of knowledge.

        Consider two accepting conversations generated using the 
        secret value $x_1$:
        \begin{align*}
          &(t_1, t_2, t_3, c_1, c_2, r_1, r_2)\\
          &(t_1, t_2, t_3, c'_1, c'_2, r'_1, r'_2)
        \end{align*}
        If both proofs are valid, then these relations holds
        (by the verification equation):
        \begin{align*}
          r_1 &= v_1 - c_1 x_1&
          r'_1 &= v_1 - c'_1 x_1
        \end{align*}
        Recovering the secret value $x_1$ (and hence demonstrating
        special soundness) requires solving this system for $x_1$.
        The equations are independent since $c_1 \neq c'_1$ by definition.
        \begin{align*}
          x_1 = \frac{r_1 - r'_1}{c'_1 - c_1}
        \end{align*}
        
  \item \textbf{Honest-verifier computational zero-knowledge}:
        there is a simulator that, when given the statement
        and the challenge $h$ as input, produces transcripts 
        computationally indistinguishable from those that
        an honest prover would produce in responding to $h$.
   
        It is straightforward to construct a simulator that, when given $h$, 
        produces accepting transcripts that are computationally indistinguishable
        from a prover-generated transcript.

        The simulator receives $h$ as input and picks a random
        exponent $w \in_R \mathbb{Z}_q$.
        The simulator sets:
        \begin{align*}
          c_1 &= h - w\\
          c_2 &= w
        \end{align*}
        The simulator then picks $v_1, v_2 \in_R \mathbb{Z}_q$, 
        and sets the responses:
        \begin{align*}
            r_1 &= v_1 &
            r_2 &= v_2
        \end{align*}
        The commitments are then:
        \begin{align*}
            t_1 &= y_1^{c_1} g_1^{v_1}&
            t_2 &= y_2^{c_1} g_2^{v_1}&
            t_3 &= y_3^{c_2} g_3^{v_2}
        \end{align*}
        Simple substitution into the verification equation confirms
        that the simulator's transcript is valid and that it is
        computationally indistinguishable from a prover-generated one.
\end{itemize}

%% file: lazy.tex
\section{Lazy Proof Verification}
\label{app:lazy}

The {\em lazy proof verification} technique
described in Section~\ref{sec:crypto-hashing-lazy}
is an optimization in which
servers to process client ciphertexts
without checking the ciphertexts' proofs of validity.
Servers only check the client proofs if they
fail to recover the slot owner's plaintext message
later in the protocol run.

Implementing lazy proof verification is
straightforward: when the slot owner submits
the plaintext message $m$ to a run of the
protocol, the slot owner constructs $m$ such that it
consists of the plaintext message $m'$ and
a cryptographic signature (using the slot owner's pseudonym
signing key) of $m'$:\footnote{%
In practice, the value $m$ is encoded 
as a {\em vector} of group elements. 
To simplify our exposition, we pretend 
in this section that $m$ consists
of a single group element.}
\begin{align*}
  m = \langle m', \textsc{sign}_\textrm{SlotOwner}(m') \rangle
\end{align*}
The slot owner then submits $m$ as the plaintext message
to the \app protocol run.
When servers reach the point in the protocol in which they 
obtain the slot owner's plaintext message $m$, they parse
$m$ into $\langle m', \sigma \rangle$.
If $\sigma$ is a valid signature on $m'$, the servers
return $m'$ to the clients as the slot owner's plaintext.
Otherwise, the servers check each of the client proofs
to expose the dishonest client.

Lazy proof verification introduces a subtle security
issue that makes it insecure with
the ElGamal-style construction (but secure with the other schemes).
If servers use lazy proof verification 
with the ElGamal-style ciphertext construction, 
a malicious client could submit a ciphertext that ``cancels out'' the 
ciphertexts of other clients, thereby violating 
the security properties of the system.

For example, consider a protocol run in which there
are three clients---two honest clients and one dishonest
client---and two honest servers with public keys
$g^a$ and $g^b$.
Using the ElGamal-style ciphertext construction, 
The two honest clients might submit the 
following ciphertext tuples:
\begin{align*}
  \langle R_1, C_1, \textsf{PoK}_1 \rangle &= \langle g^r, mg^{(a+b)r}, \textsf{PoK}_1 \rangle\\
  \langle R_2, C_2, \textsf{PoK}_2 \rangle &= \langle g^s, g^{(a+b)s}, \textsf{PoK}_2 \rangle
\end{align*}
Given these two correct client ciphertexts, the dishonest
client makes a guess that client 1 is the slot owner.
To confirm this guess, the dishonest client constructs
a client ciphertext with an {\em invalid} proof of
knowledge:
\begin{align*}
  \langle R_3, C_3, \textsf{PoK}_3 \rangle &= \langle R_2^{-1}, C_2^{-1}, \textsf{PoK}_3 \rangle\\
    &= \langle g^{-s}, g^{-(a+b)s}, \textsf{PoK}_3 \rangle
\end{align*}
The dishonest client can construct this ciphertext
because it is easy to invert elements of the groups we use.
The proof of knowledge $\textsf{PoK}_3$ is invalid
because dishonest client 3 does not know the secret
exponent $-s$ and therefore cannot (by the special soundness
property) cannot construct a valid proof of knowledge.

If the servers accept these three ciphertexts without
verifying the accompanying proofs of knowledge, 
the servers will generate their server ciphertext
elements in response to the ciphertexts
$\langle C_1, C_2, C_3 \rangle$:
\begin{align*}
  S_1 = g^{-a(r+s-s)} & & S_2 = g^{-b(r+s-s)}
\end{align*}
The servers then compute the slot owner's plaintext
message $m^*$ as:
\begin{align*}
  m^* &= (C_1 C_2 C_3) (S_1 S_2)\\
    &= (mg^{(a+b)r} g^{(a+b)s} g^{-(a+b)s})(g^{-a(r+s-s)} g^{-b(r+s-s)})\\
    &= (mg^{(a+b)r} g^{(a+b)s} g^{-(a+b)s})(g^{-ar} g^{-br})\\
    &= mg^{ar+br} g^{as+bs} g^{-as-bs} g^{-ar-br}\\
    &= m\\
    &= \langle m', \textsc{sign}_\textrm{SlotOwner}(m') \rangle
\end{align*}
Since the servers recover the slot owner's original signed
plaintext message $m'$, the servers will not suspect
any malicious behavior on the part of the dishonest client.

The dishonest client, however, has 
learned that client 1 is the anonymous slot owner.
The dishonest client knows this because his ciphertext
message $C_3$ effectively cancels out the effect of
honest client $C_2$'s ciphertext 
(because $C_3$ is the inverse of $C_2$).
If instead client 2 was the anonymous slot owner, then the client 3's
malicious ciphertext would have cancelled out the plaintext
message encoded in client 2's ciphertext, 
and the servers would have concluded that $m^* = 1$.
Since the servers successfully recovered the slot owner's plaintext message,
client 3 correctly concludes that client 1 was the slot owner.

The sketch above demonstrates why lazy proof verification
is insecure when used with ElGamal-style ciphertexts.
We now argue that lazy proof verification is 
secure in the hashing-generator construction.
(The security argument for the pairing-based scheme
has a similar structure.)

In the hashing-generator construction, every client $i$
and every server $j$ share a secret $r_{ij}$.
Consider the attack scenario above, with two
honest clients and one dishonest client.
The honest clients' ciphertexts in messaging
round $\ell$ will be:
\begin{align*}
  C_1 &= m g_\ell^{r_{11}+r_{12}}\\
  C_2 &= g_\ell^{r_{21}+r_{22}}
\end{align*}

The dishonest client can try a trick 
similar to the one demonstrated before---%
she constructs a ciphertext that 
contains her own ciphertext multiplied by the
inverse of $C_2$:
\begin{align*}
  C_3 &= C'_3 C_2^{-1} \\
      &= g_\ell^{r_{31} + r_{32}} g_\ell^{-r_{21}-r_{22}}
\end{align*}
The servers then construct their ciphertexts:
\begin{align*}
  S_1 &= g_\ell^{-r_{11}-r_{21}-r_{31}}\\
  S_2 &= g_\ell^{-r_{12}-r_{22}-r_{32}}
\end{align*}

When the servers take the product of
all of the client and server ciphertexts
to reveal the slot owner's message $m^*$, the
result is corrupted:
\begin{align*}
  m^* =& (C_1 C_2 C_3)(S_1 S_2)\\
      =&(m g_\ell^{r_{11}+r_{12}} g_\ell^{r_{21}+r_{22}} g_\ell^{r_{31} + r_{32} -r_{21}-r_{22}})\\
       &(g_\ell^{-r_{11}-r_{21}-r_{31}} g_\ell^{-r_{12} -r_{22} - r_{32}})\\
      =&m g_\ell^{r_{11}+r_{12}} g_\ell^{-r_{11}-r_{21} -r_{12} -r_{22}}\\
      =&m g_\ell^{-r_{21} -r_{22}}
\end{align*}
In this case, the servers will notice the corruption, will
verify the client proofs of knowledge, and will
expose client 3 as dishonest.
In addition, since the slot owner's plaintext message
would also be corrupted if the dishonest client
chose to attack client 1's ciphertext, the dishonest
client does not learn which honest client is the slot owner.

This attack fails in the general case as well.
In the hashing-generator construction,
each honest server produces the same server
ciphertext element in messaging round $\ell$
irrespective of the ciphertexts that the clients
submit in that round.
(This is in contrast with the ElGamal-style
construction, in which server ciphertexts
depend on the client ciphertexts.)
A dishonest client therefore does not learn 
{\em any} additional information from an honest 
client or server by submitting an incorrectly
formed ciphertext.
Lazy proof verification is then secure
when using the hashing-generator 
ciphertext construction.